\documentclass[longauth,ceqn]{aa}  

\usepackage{graphicx}
\usepackage{txfonts}
\usepackage{amsmath}
\usepackage{amssymb}

\usepackage{orcidlink}

\usepackage{caption}
\usepackage{subcaption}
\usepackage{makecell}

\usepackage{hyperref}
\usepackage{natbib}

\hypersetup{
  colorlinks   = true, 
  urlcolor     = blue, 
  linkcolor    = blue, 
  citecolor   = blue 
}
\begin{document}

   \title{Confirmation of four hot Jupiters detected by TESS using follow-up spectroscopy from MaHPS at Wendelstein together with NEID and TRES}

    \author{
    Juliana Ehrhardt \orcidlink{0009-0003-9433-043X}\inst{\ref{inst1}, \ref{inst2}}
    \and Luis Thomas \orcidlink{0009-0006-1571-0306} \inst{\ref{inst1}, \ref{inst2}}
    \and Hanna Kellermann \orcidlink{0009-0006-3527-0424} \inst{\ref{inst1}} 
    \and Christine Freitag \inst{\ref{inst1}}
    \and Frank Grupp \inst{\ref{inst1}, \ref{inst2}}
    \and Samuel W.\ Yee \orcidlink{0000-0001-7961-3907} \inst{\ref{inst3}, \ref{inst4}, \ref{inst5}}
    \and Joshua N.\ Winn \orcidlink{0000-0002-4265-047X} \inst{\ref{inst4}}
    \and Joel D.\ Hartman \orcidlink{0000-0001-8732-6166} \inst{\ref{inst4}}
    \and Karen A.\ Collins \orcidlink{0000-0001-6588-9574} \inst{\ref{inst8}}
    \and Cristilyn N.\ Watkins \orcidlink{0000-0001-8621-6731} \inst{\ref{inst8}}
    \and Keivan G.\ Stassun \orcidlink{0000-0002-3481-9052} \inst{\ref{inst9}}
    \and Paul Benni \orcidlink{0000-0001-6981-8722} \inst{\ref{inst10}}
    \and Allyson Bieryla \orcidlink{0000-0001-6637-5401} \inst{\ref{inst12a}}
    \and Kylee Carden \orcidlink{0009-0008-9182-7471} \inst{\ref{inst13}, \ref{inst14}, \ref{inst15}}
    \and Jacek Checinski \orcidlink{0000-0001-6352-730X} \inst{\ref{inst38}}
    \and Dmitry V.\ Cheryasov \orcidlink{0009-0003-4203-9667} \inst{\ref{inst16}}
    \and Brendan Diamond \inst{\ref{inst17}}
    \and Nicholas Dowling \inst{\ref{inst1}}
    \and Courtney D. Dressing \orcidlink{0000-0001-8189-0233} \inst{\ref{inst18}}
    \and Emma Esparza-Borges \orcidlink{0000-0002-2341-3233} \inst{\ref{inst11}, \ref{inst12}}
    \and Phil Evans \orcidlink{0000-0002-5674-2404} \inst{\ref{inst19}}
    \and Raquel For\'{e}s-Toribio \orcidlink{0000-0002-6482-2180} \inst{\ref{inst20}, \ref{inst21}}
    \and Akihiko Fukui \orcidlink{0000-0002-4909-5763} \inst{\ref{inst22}, \ref{inst11}}
    \and Steven Giacalone \orcidlink{0000-0002-8965-3969} \inst{\ref{inst24}, \ref{inst25}}
    \and Eric Girardin \inst{\ref{inst26}}
    \and Robert F. Goeke \inst{\ref{inst26a}}
    \and Claus Goessl \inst{\ref{inst1}}
    \and Yuya Hayashi \orcidlink{0000-0001-8877-0242} \inst{\ref{inst27}}
    \and Ulrich Hopp \inst{\ref{inst1}, \ref{inst2}}
    \and Jon M. Jenkins \orcidlink{0000-0002-4715-9460} \inst{\ref{inst27a}}
    \and Isa Khan \inst{\ref{inst28}}
    \and Didier Laloum \inst{\ref{inst29}}
    \and Adam Lark \inst{\ref{inst28}}
    \and David W. Latham \orcidlink{0000-0001-9911-7388} \inst{\ref{inst12a}}
    \and Jerome de Leon \orcidlink{0000-0002-6424-3410} \inst{\ref{inst22}, \ref{inst11}}
    \and Alessandro Marchini \orcidlink{0000-0003-3779-6762} \inst{\ref{inst31}}
    \and Bob Massey \orcidlink{0000-0001-8879-7138} \inst{\ref{inst20}}
    \and Jose A. Mu\~noz \orcidlink{0000-0001-9833-2959} \inst{\ref{inst20}, \ref{inst21}}
    \and Felipe Murgas \orcidlink{0000-0001-9087-1245} \inst{\ref{inst11}, \ref{inst12}}
    \and Norio Narita \orcidlink{0000-0001-8511-2981} \inst{\ref{inst22}, \ref{inst36}, \ref{inst11}}
    \and Enric Palle \orcidlink{0000-0003-0987-1593} \inst{\ref{inst11}, \ref{inst12}}
    \and Riccardo Papini \orcidlink{0009-0006-9361-9153}\inst{\ref{inst37}}
    \and Hannu Parviainen \orcidlink{0000-0001-5519-1391} \inst{\ref{inst12}, \ref{inst11}}
    \and Jan-Niklas Pippert \orcidlink{0009-0006-9461-002X} \inst{\ref{inst1}, \ref{inst2}}
    \and Adam Popowicz \orcidlink{0000-0003-3184-5228} \inst{\ref{inst38}}
    \and Tyler Pritchard \inst{\ref{inst38a}}
    \and Samuel N. Quinn \orcidlink{0000-0002-8964-8377} \inst{\ref{inst12a}}
    \and Manfred Raetz \inst{\ref{inst39}}
    \and Christoph Ries \inst{\ref{inst1}}
    \and Arno Riffeser \inst{\ref{inst1}}
    \and Arjun B.\ Savel \orcidlink{0000-0002-2454-768X} \inst{\ref{inst40}}
    \and Sara Seager \orcidlink{0000-0002-6892-6948} \inst{\ref{inst26a}, \ref{inst40aa}, \ref{inst40ab}}
    \and Michael Schmidt \inst{\ref{inst1}}
    \and Stephanie Striegel \orcidlink{0009-0008-5145-0446} \inst{\ref{inst40a}, \ref{inst27a}}
    \and Gregor Srdoc \inst{\ref{inst41}}
    \and Chris Stockdale \orcidlink{0000-0003-2163-1437} \inst{\ref{inst42}}
    \and Gaia Verna \orcidlink{0000-0001-5916-9028} \inst{\ref{inst31}}
    \and David Watanabe \orcidlink{0000-0002-3555-8464} \inst{\ref{inst42a}}
    \and Carl Ziegler \inst{\ref{inst43}}
    \and Raphael Zöller \orcidlink{0000-0002-0938-5686} \inst{\ref{inst1},\ref{inst2}}
    }
    
    \institute{
    Universit\"ats-Sternwarte M\"unchen, Fakult\"at für Physik,  Ludwig-Maximilians-Universit\"at M\"unchen, Scheinerstr. 1, D-81679 M\"unchen, Germany\label{inst1}
    \and Max Planck Institute for Extraterrestrial Physics, Giessenbachstrasse 1, 85748 Garching, Germany\label{inst2}
    \and Center for Astrophysics \textbar \ Harvard \& Smithsonian, 60 Garden St, Cambridge, MA 02138, USA\label{inst3}
    \and 
    Department of Astrophysical Sciences, Princeton University, 4 Ivy Lane, Princeton, NJ 08544, USA\label{inst4} 
    \and 51 Pegasi b Fellow\label{inst5}
    \and Center for Astrophysics \textbar \ Harvard \& Smithsonian, 60 Garden Street, Cambridge, MA 02138, USA \label{inst8}
    \and Department of Physics and Astronomy, Vanderbilt University, Nashville, TN 37235, USA \label{inst9}
    \and Acton Sky Portal private observatory, Acton, MA, USA \label{inst10}
    \and Instituto de Astrof\'{i}sica de Canarias (IAC), 38205 La Laguna, Tenerife, Spain \label{inst11}
    \and Departamento de Astrofísica, Universidad de La Laguna (ULL), 38206, La Laguna, Tenerife, Spain \label{inst12}
    \and Center for Astrophysics ${\rm \mid}$ Harvard {\rm \&} Smithsonian, 60 Garden Street, Cambridge, MA 02138, USA \label{inst12a}
    \and Department of Astronomy, The Ohio State University, Columbus, OH 43210, USA\label{inst13}
    \and Thüringer Landessternwarte Tautenburg, 07778 Tautenburg, Germany \label{inst14}
    \and Department of Physics, Massachusetts Institute of Technology, Cambridge, MA 02139, USA \label{inst15}
    \and Sternberg Astronomical Institute, Lomonosov Moscow State University, Universitetsky prospekt, 13, Moscow 119992, Russia \label{inst16}
    \and Howard Community College, 10901 Little Patuxent Pkwy, Columbia, MD 21044, USA \label{inst17}
    \and Department of Astronomy, University of California Berkeley, Berkeley, CA 94720, USA \label{inst18}
    \and Phil Evans, El Sauce Observatory, Coquimbo Province, Chile \label{inst19}
    \and Departamento de Astronom\'{\i}a y Astrof\'{\i}sica, Universidad de Valencia, E-46100 Burjassot, Valencia, Spain \label{inst20}
    \and Observatorio Astron\'omico, Universidad de Valencia, E-46980 Paterna, Valencia, Spain \label{inst21}
    \and Komaba Institute for Science, The University of Tokyo, 3-8-1 Komaba, Meguro, Tokyo 153-8902, Japan \label{inst22}
    \and NSF Astronomy and Astrophysics Postdoctoral Fellow \label{inst24}
    \and Department of Astronomy, California Institute of Technology, Pasadena, CA 91125, USA \label{inst25}
    \and Grand Pra Observatory, Switzerland \label{inst26}
    \and Department of Physics and Kavli Institute for Astrophysics and Space Research, Massachusetts Institute of Technology, Cambridge, MA 02139, USA \label{inst26a}
    \and Department of Multi-Disciplinary Sciences, Graduate School of Arts and Sciences, The University of Tokyo, 3-8-1 Komaba, Meguro, Tokyo 153-8902, Japan \label{inst27}
    \and NASA Ames Research Center, Moffett Field, CA 94035, USA \label{inst27a}
    \and Hamilton College, 198 College Hill Rd, Clinton, NY 13323 \label{inst28}
    \and Société Astronomique de France, 3 Rue Beethoven, 75016 Paris, France \label{inst29}
    \and Astronomical Observatory, University of Siena, 53100 Siena, Italy \label{inst31}
    \and Villa '39 Observatory, Landers, CA 92285, USA \label{inst32}
    \and Astrobiology Center, 2-21-1 Osawa, Mitaka, Tokyo 181-8588, Japan \label{inst36}
    \and Wild Boar Remote Observatory, San Casciano in val di Pesa, Firenze, 50026 Italy \label{inst37}
    \and Silesian University of Technology, Akademicka 2A, 44-100, Gliwice, Poland \label{inst38}
    \and NASA Goddard Space Flight Center, 8800 Greenbelt Road, Greenbelt, MD 20771, USA \label{inst38a}
    \and Privat Observatory Herges-Hallenberg, Steinbach-Hallenberg, Germany \label{inst39}
    \and Department of Astronomy, University of Maryland, College Park, College Park, MD 20742 USA \label{inst40}
    \and Department of Earth, Atmospheric and Planetary Sciences, Massachusetts Institute of Technology, Cambridge, MA 02139, USA \label{inst40aa}
    \and Department of Aeronautics and Astronautics, MIT, 77 Massachusetts Avenue, Cambridge, MA 02139, USA \label{inst40ab}
    \and SETI Institute, Mountain View, CA 94043 USA/NASA \label{inst40a}
    \and Kotizarovci Observatory, Sarsoni 90, 51216 Viskovo, Croatia \label{inst41}
    \and Hazelwood Observatory, Australia \label{inst42}
    \and Planetary Discoveries in Fredericksburg, VA 22405, USA \label{inst42a}
    \and Department of Physics, Engineering and Astronomy, Stephen F. Austin State University, 1936 North St, Nacogdoches, TX 75962, USA \label{inst43}
    }

   \date{Received Month Date, 2024; accepted month date, 2024}

 
\abstract{
We report the confirmation and characterization of \ four hot Jupiter-type exoplanets initially
detected by TESS:\ TOI-1295 b, TOI-2580 b, TOI-6016 b, and TOI-6130 b. Using observations with the high-resolution echelle spectrograph MaHPS on the 2.1m telescope at Wendelstein Observatory, together with NEID at Kitt Peak National Observatory and TRES at the Fred Lawrence Whipple Observatory, we confirmed the planetary nature of these four planet candidates. We also performed precise mass measurements. All four planets are found to be hot Jupiters with orbital periods between 2.4 and 4.0~days. The sizes of these planets range from 1.29 to 1.64 Jupiter radii, while their masses range from 0.6 to 1.5 Jupiter masses. Additionally, we investigated whether there are signs of other planets in the systems but have found none. Lastly, we compared the radii of our four objects to the results of an empirical study of radius inflation and see that all four demonstrate a good fit  with the current models. These four planets belong to the first array of planets confirmed with MaHPS data, supporting the ability of the spectrograph to detect planets around fainter stars as faint as $V=12$.}

\titlerunning{Confirmation of Four Hot Jupiters with MaHPS, NEID, and TRES}
\authorrunning{Ehrhardt et al.}
   \maketitle
%
\section{Introduction}
The hot Jupiter 51 Pegasi b \citep{Mayor1995} was the first exoplanet to be detected around a Sun-like star. To date,  more than 5600 exoplanets have been confirmed  in the {\tt NASA Exoplanet Archive}\footnote{\href{https://exoplanetarchive.ipac.caltech.edu/index.html}{https://exoplanetarchive.ipac.caltech.edu/index.html}} with about 10\% of the detected exoplanets being hot Jupiters. Hot Jupiters are usually defined as planets having a mass exceeding about 0.25\,$\text{M}_J$ and a short orbital period ($P \leq 10$~days), indicating they are in close proximity to their host stars. 

Despite hot Jupiters accounting for most of the early planet detections, they are not a common occurrence in the universe, making up only $\sim 1\%$ of solar-type stars.
They are even more rare around smaller (M-type) dwarf stars \citep[e.g.,  ][]{Dawson2018,Zhou2019}.
Due to their orbital properties, hot Jupiters are the most accessible to observe and characterize. Despite their low occurrence rate, about 10\% of the currently confirmed exoplanets are classed as hot Jupiters \citep{Schulte2024}.
The detection of hot Jupiters has put common formation theories derived from the Solar System into question. Even though the formation mechanism for hot Jupiters remains a subject of debate, it is widely assumed that they form in one or a combination of the following processes \citep[see][]{Dawson2018}.
\begin{enumerate}
\item In situ formation: The planet would form at its current position via core accretion or disk instability.
\item  Disk migration: The planet perturbs surrounding gas onto horseshoe orbits and deflects gas over large distances. Via this mechanism, it can clear a gap and migrate inward (Type II migration).
We do not consider type I migration here, as we focus on the formation of hot Jupiters; whereas the type 1 mechanism is crucial for low-mass planets, specifically \citep{Bitsch2019}.
\item  High-eccentricity tidal migration: This process is assumed to occur in two steps. First, the planet's orbital angular momentum is reduced. This is assumed to happen due to another planet or a star, which disturbs the hot Jupiter's orbit to an elliptical shape. Possible mechanisms at work here are planet-planet scattering (with nearby planets) or secular interactions (with more distant planets). 
After this process, the exoplanet undergoes very close encounters with the star (at its periapsis).
Furthermore, tidal interactions with the star cause the exoplanet to periodically change its form, making it lose orbital energy and slowly circularize into a short-period orbit. As the distance to the perturber (which initially caused the increase in the ellipticity of the planet) grows during this process, the final hot Jupiter planet eventually ends up unperturbed by these companions. 
\end{enumerate}
Observing hot Jupiters and finding out more about their properties and the systems they are part of could offer clues on their origins and formation mechanisms. Important observables to distinguish between the different formation models are the eccentricity and obliquity.
The existence of a wide ranges of eccentricities and obliquities seems to imply more turbulent dynamical formation histories. Additional constraints are offered by the presence and properties of planetary companions.
\\
Both in situ formation and disk migration would allow for other planets to exist in close proximity to the hot Jupiter \citep{Lee:2002, Hansen:2013, Batygin:2016, Boley:2016}. Compared to in situ formation, for disk migration, this would lead to orbits that are preferably in mean motion resonance although some planets may be able to escape this resonance \citep{Goldreich:2014}. On the other hand, a high-eccentricity tidal migration would wipe out most inner planets during the migration process of the hot Jupiter progenitor leading to an isolated planet \citep{Mustill:2015,Zink2023}. In this case, a stellar or giant planet companion is needed to trigger the migration. For planet-planet scattering, the giant companion can be ejected from the system \citep{Chatterjee:2008}; however, for Lidov-Kozai cycling \citep{Fabrycky:2007,Naoz:2016} or secular scattering \citep[e.g.,   ][]{Wu:2011}, a long-period companion is expected to be present in the system. 
\\
Studies of the nature of hot Jupiter systems have found that stellar companions close enough to incite high-eccentricity migration are rare \citep{Ngo:2016}. On the other hand, most hot Jupiter systems seem to have an outer giant planet companion that could be responsible for the migration of the hot Jupiter \citep{Knutson:2014,Bryan:2016,Zink2023}. Together with the lack of detections of nearby companions to hot Jupiters from transit timing variation (TTV) analyses \citep{Steffen2012,Huang:2016}, this would suggest high-eccentricity migration as the dominant mechanism for hot Jupiter formation. However, more recently, some hot Jupiter systems with close-in small planet companions have been detected \citep[for example  ][]{Canas:2019,Huang:2020,Hord:2022} and an analysis of the full Kepler baseline found TTVs for at least 12~\% $\pm$ 6~\% of hot Jupiters \citep{Wu2023}.
\\
Additionally, the search for correlations between planetary and stellar parameters (examples of the latter would be its spin period, mass, radius, metallicity, age, and spectral type) can provide insights into the complete evolutionary pathway of planetary systems \citep{vonBraun2017}.
\\
Another important open question regarding hot Jupiters is the so-called radius anomaly. The theoretically expected radii of hot Jupiters have been derived starting from a standard theory of giant planets \citep{2002Hubbard}, as well as other effects, such as the evolution of the planetary radii with stellar radiation \citep[for example ][]{Fortney2010}.
However, in a large number of detected exoplanets, the observed radii of the hot Jupiters are found to be larger than theoretically expected \citep[][and references therein]{Demory2011}.
Currently, there are several theories for the processes responsible for this inflation \citep[for example ][]{Fortney2010}, however, no definitive conclusion has been made yet \citep{Thorngren2021}. Understanding the inflation mechanisms will be crucial to gaining information about the interiors of giant exoplanets,
as well as their evolution and formation \citep[for example ][]{Thorngren2021}.
This paper aims to contribute to the above-listed issues on the understanding of hot Jupiters.
\\
In this work, four hot Jupiter planet candidates from TESS:\ TOI-1295, TOI-2580, TOI-6016, and TOI-6130, are confirmed as exoplanets. Several properties, such as their radii, masses, etc., have been determined using joint fits with the Python package {\tt juliet} of the radial velocities from spectroscopic measurements and photometric transit data (Sects. \ref{Sec:PhotObservations}-\ref{Sec:Analysis}).
The resulting values from the fits are discussed (Sect. \ref{Sec:Discussion}) in the context of open questions about hot Jupiter formation and evolution, such as inflation or the presence of companion planets.

\section{Photometric observations} \label{Sec:PhotObservations}

\subsection{TESS photometry} \label{Sec:photobsTESS}

 Transiting Exoplanet Survey Satellite (TESS) was developed to detect exoplanets around bright and nearby stars using the transit detection method \citep{Ricker2015}.
For this purpose, TESS observes the entire sky by separating it into observing sectors. These sectors are observed with full-frame images (FFIs) with a 30 minute cadence (prime mission), 10 minute observation timespan (extended mission), and 200~second cadence (second extended mission) over a time span of about 27~days \citep{jenkinsSPOC2016}.
TESS uses four cameras, each with a field of view (FoV) of 24$^\circ$ x 24$^\circ$. These four cameras are aligned vertically in a way that during each sector a field of 24$^\circ$ x 96$^\circ$ is observed. Additionally, postage stamp images are downloaded for objects on a specific target list with a 2 minute cadence. A pixel in such a FFI or postage stamp image has a size of 21.1”. 
The observing bandpass of TESS reaches from about 600~nm to 1000~nm \citep{Sullivan2015}.
More comprehensive information on TESS can be found in the TESS handbook \footnote{\href{https://archive.stsci.edu/files/live/sites/mast/files/home/missions-and-data/active-missions/tess/_documents/TESS_Instrument_Handbook_v0.1.pdf}{https://archive.stsci.edu/files/live/sites/mast/files/home/missions-and-data/active-missions/tess/\_documents/TESS\_Instrument\_Handbook\_v0.1.pdf}}.
The image data were reduced and analyzed by the Science Processing Operations Center (SPOC) at NASA Ames Research Center.
\\
The objects in the TOI \footnote{TOI: TESS Object of Interest} catalog should be re-observed photometrically and/or spectroscopically to confirm their nature as exoplanets, constrain their parameters, and gain additional information (for example about their atmospheres). Supplementary observations might be needed as the limited information about the host star can restrict the precision of the exoplanet parameters.
\\
The photometric light curves from the TESS data are downloaded for each sector selecting the data reduced with the 'TESS-SPOC'-pipeline \citep{Caldwell2020} and using the Python package {\tt lightkurve} \citep{LightkurvePythonPackage}. We have used the {\tt pdcsap}-flux-values, as they have been corrected for systematics.
\\
\subsubsection{Transit observations}

\textbf{TOI-1295} (TIC 219852584) was observed in the full-frame images from TESS in Sectors 14-26 (both included) with a 30-minute cadence, in Sectors 40, 41, and 47 - 55 with a 600~second cadence, and in Sectors 56, 57, 59, and 60 with a 200~second cadence. The planetary details reported on the ExoFOP website are: transit depth is 7960 $\pm$ 30~ppm and an average period of 3.197~days.
\\
\textbf{TOI-2580} (TIC 102713734) was observed in the full-frame images from TESS in Sector 19 with a 30 minute cadence and in Sector 59 with a 200~s cadence, additionally to observations with a 2 minute cadence. It was first found as a CTOI \footnote{CTOI: community TOI} \citep[see][]{Olmschenk_2021}. Afterward, the planetary details were refined and reported on the ExoFOP website as transit depth is 9960 $\pm$ 0.9~ppm and an average period of 3.398~days.
\\
\textbf{TOI-6016} (TIC 327369524) was observed in the full-frame images from TESS in Sectors 17, 18, and 24 with a 30 minute cadence, as well as in Sectors 57 and 58 both with a 200~s and a 120~s cadence. It was first found as a CTOI \citep[see][]{Olmschenk_2021}. Afterward, the planetary details were refined and reported on the ExoFOP website as transit depth is 8925 $\pm$ 178~ppm and an average period of 4.024~days.
The SPOC conducted a transit search of Sector 57 on 1 November 2022 using an adaptive, noise-compensating matched filter \citep{Jenkins2002, Jenkins2010, Jenkins2020}, producing a TCE for which an initial limb-darkened transit model was fitted \citep{Li2019}. Furthermore, a suite of diagnostic tests were conducted to help make or break the planetary nature of the signal \citep{Twicken2018}. The TESS Science Office (TSO) reviewed the vetting information and issued an alert for TOI-6016.01 on 15 December 2022 \citep{guerrer2021}. According to the difference image centroiding tests, the host star is located within 0.27 $\pm$ 2.5 $\farcs$ of the transit signal source of TOI-6016.
\\
\textbf{TOI-6130} (TIC 210083929) was observed in the full-frame images from TESS in Sector 56 with a 200~s cadence. The planetary details were reported on the ExoFOP website as transit depth is 13405 $\pm$ 305~ppm and an average period of 2.393~days.

\subsubsection{SPOC difference imaging analysis}

Since the PSF of the TESS images is slightly undersampled with a FWHM of $1\farcs2$ (similar to Kepler's and to Hubble's WFC3 PSF scales), this permits the SPOC difference image centroiding analyses to constrain the location of transit sources to typically within $\pm 2\farcs5$, especially for strong transit signals. This is the case for all four planets announced in this paper.
The TIC offset in the TESS SPOC DV reports constrains the distance of the target star from the transit source location.
\\
\textbf{TOI-1295:} The difference image analysis for sector 23 constrained the host star to within 0.564 $\pm$ 2.45 $\farcs$ of the transit source location.
\\
\textbf{TOI-2580:} The difference image analysis for sector 59 constrained the host star to within 0.340 $\pm$ 2.4 $\farcs$ of the transit source location.
\\
\textbf{TOI-6016:} The difference image analysis for sector 58 constrained the host star to within 0.274 $\pm$ 2.5 $\farcs$ of the transit source location, respectively.
\\
\textbf{TOI-6130:} The difference image analysis for sector 56 constrained the host star to within 0.226 $\pm$ 2.5 $\farcs$ of the transit source location.
\\
Since the SPOC difference image analysis is sensitive to background transit sources out to the edge of the postage stamp, these results complement the high-resolution imaging data (see Sect. \ref{Sec:HiAngResObs}), which have much smaller FOVs and are available as soon as the TESS data are released to the community through MAST. 

\subsection{Ground-based time domain photometry} \label{Sec:photobsGroundBased}

For all four objects, several full transits were observed with ground-based telescopes. These observations can be used to improve the accuracy of the light curve parameters; for example, refining the timing of the transit event (both its midpoint and the duration). They may also put better constraints on the shape of the transit curve. Additionally, possible variations in the transit timing could be detected, signaling a potential multi-planet system.
\\
Due to the low spatial resolution of TESS ($\sim$ 21" per pixel), it is often the case that there are numerous stars inside one pixel. Follow-up observations with a higher spatial resolution are needed to rule out, that the transit signal is from a neighbor star.
\\
The data for the ground-based observations were prepared using the code {\tt AstroImageJ} \citep{Collins2017} as time-series data with detrended normalized relative flux values and the relative flux uncertainty. 
Only the observations of TOI-6130 obtained by the MUSCAT2 team utilised a dedicated MuSCAT2 pipeline. In \citet{Parviainen2020}, the performance of the data reduction and extraction of the photometry are described.

\subsubsection*{TOI-1295} \label{Sec:1295_GroundBasedPHot}

For TOI-1295, we observed nine full transits and one egress. The fitted light curves are shown online (see Sec. \ref{Sec:DataAvailability}).\\
\begin{table}[htp]
    \centering
    \caption{Overview of photometric time series observations of TOI-1295}
    \begin{tabular}{c|c|c|c|c}
Date & Tel. size & Band & Image scale & Apertures\\
\hline \hline
2019-11-07 &  0.36\,m & $r'$ & $0\farcs712$ & $2\farcs2$ \\
2020-04-01 &  0.4\,m & $g'$ & $0\farcs73$ & $4\farcs4$ \\
2020-04-17 & 0.3\,m & R & $1\farcs2$ & $15\farcs8$ \\
2020-05-03  & 0.3\,m & R & $1\farcs2$ & $10\farcs1$ \\
2020-05-03  & 0.2\,m & CBB & $0\farcs69$ & $6\farcs2$ \\
2020-11-11 & 0.4\,m & $r'$ & $0\farcs635$ & $5\farcs1$ \\
2021-05-12  & 0.5\,m & B, $i'$ & $0\farcs54$ & $3\farcs8$ \\
2021-06-13 & 0.3\,m & B & $0\farcs685$ & $8\farcs2$ \\
2022-04-16 & 0.28\,m & clear & $1\farcs02$ & $9\farcs3$ \\
2023-04-09 & 0.4\,m & $g'$ & $0\farcs264$ & $2\farcs7$ \\
\end{tabular}
\end{table}
\\
We observed a partial transit window of TOI-1295.01 in Sloan $r'$ on UTC 7 November 2019 from Howard Community College in Columbia, Maryland. The 0.36\,m telescope is equipped with a SBIG STXL-6303 detector with an image scale of $0\farcs712$~pixel$^{-1}$. The differential photometric data were extracted using {\tt AstroImageJ}. We used circular photometric apertures with a radius of $2\farcs2$, which excluded the flux from the nearest known neighbor in the Gaia DR3 catalog (Gaia DR3 1636693568423807360), which is $\sim37\farcs1$ northeast of TOI-1295.
\\
We observed a full transit window of TOI-1295.01 in Sloan $g'$ on UTC 1 April 2020 from the Grand-Pra Observatory in Valais Sion, Switzerland. The 0.4\,m RCO telescope is equipped with a FLI 4710 detector with an image scale of $0\farcs73$~pixel$^{-1}$, resulting in a $12.9\farcm2\times12.55\farcm2$ FoV. The differential photometric data were extracted using {\tt AstroImageJ}. We used circular photometric apertures with a radius of $4\farcs4$, which excluded the flux from the nearest known neighbor in the Gaia DR3 catalog.
\\
We observed two full transit windows of TOI-1295.01 in Baader R 610\,nm longpass band on UTC 17 April 2020 and 3 May 2020 from the Kotizarovci Private Observatory 0.3\,m telescope near Viskovo, Croatia. The telescope is equipped with a $765\times510$ pixel SBIG ST7XME detector, which has an image scale of $1\farcs2$ per pixel, resulting in a $15\arcmin\times10\arcmin$ FoV. The images were calibrated and the differential photometric data were extracted using {\tt AstroImageJ}, with circular photometric apertures having a radius of $15\farcs8$ and $10\farcs1$, respectively. They have also excluded the flux from the nearest known neighbor in the Gaia DR3 catalog.
\\
We observed a full transit window of TOI-1295.01 in CBB on UTC 3 May 2020 from the Private observatory of the Mount 0.2\,m telescope in Saint-Pierre-du-Mont, France. The telescope is equipped with a $3326\times2504$ pixel Atik 383 detector, which has an image scale of $0\farcs69$ per pixel, resulting in a $38\arcmin\times29\arcmin$ FoV. The differential photometric data were extracted using {\tt AstroImageJ}. We used circular photometric apertures with a radius of $6\farcs2$, which excluded the flux from the nearest known neighbor in the Gaia DR3 catalog.
\\
We observed a full transit window of TOI-1295.01 in alternating B and Sloan $i'$ bands on UTC 12 May 2021 from the Observatorio Astronòmic de la Universitat de València (OAUV-T50), located near Valencia, Spain. The OAUV-T50 0.5\,m telescope is equipped with a FLI ProLine detector, which has an image scale of $0\farcs54$~pixel$^{-1}$, resulting in a $37\arcmin\times37\arcmin$ FoV. The differential photometric data were extracted using {\tt AstroImageJ}. We used circular photometric apertures with a radius of $3\farcs8$, which excluded the flux from the nearest known neighbor in the Gaia DR3 catalog.
\\
We observed a full transit window of TOI-1295.01 in B on UTC 13 June 2021. The Silesian University of Technology (SUTO) 0.3\,m telescope is located near Otivar, Spain. The telescope is equipped with a $4656\times3520$ pixel ASI ZWO 1600MM detector, which has an image scale of 0$\farcs$685~pixel$^{-1}$, resulting in a $53\arcmin\times42\arcmin$ FoV. The differential photometric data were extracted using {\tt AstroImageJ}. We used circular photometric apertures with a radius of $8\farcs2$, which excluded the flux from the nearest known neighbor in the Gaia DR3 catalog.
\\
We observed a full transit window of TOI-1295.01 in a clear band on UTC 16 April 2022 from the Privat Observatory Herges-Hallenberg 0.28\,m telescope near Steinbach-Hallenberg, Germany. The telescope is equipped with a Moravian Instrument G2-1600 detector, which  has an image scale of $1\farcs$02~pixel$^{-1}$, resulting an a $27\arcmin\times41\arcmin$ FoV. The differential photometric data were extracted using {\tt AstroImageJ}. We used circular photometric apertures with a radius of $9.3\farcs$, which excluded the flux from the nearest known neighbor in the Gaia DR3 catalog.
\\
We observed two full transit windows of TOI-1295.01 in  Sloan $r'$ and  Sloan $g'$ on UTC 11 November 2020 and 19 April 2023 from the Wendelstein Observatory of the LMU
Munich, Germany. The UTC 11 November 2020 observation was performed when the  0.4\,m telescope was equipped with a SBIG STX-16803 camera that had an image scale of $0\farcs635$. The 9 April 2023 observation was performed with a QHY 600M camera that has an image scale of $0\farcs264$, resulting in a $42\arcmin\times28\arcmin$ FoV. Image data from all observations were calibrated and photometric data were extracted using {\tt AstroImageJ}. We used circular photometric apertures with a radius of $5\farcs1$ for the Sloan $r'$  band observation and $2\farcs7$ for the Sloan $g'$ band observation.
The light curve data are available on the {\tt EXOFOP-TESS} website\footnote{\href{https://exofop.ipac.caltech.edu/tess/target.php?id=219852584}{https://exofop.ipac.caltech.edu/tess/target.php?id=219852584}} and are included in the global modeling described in Sect. \ref{Sec:AnalysisJointFit}. 

\subsubsection*{TOI-2580} \label{Sec:2580_GroundBasedPHot}



There were two (nearly-)full transits observed of TOI-2580 and the fitted lightcurves are shown online (see Sect. \ref{Sec:DataAvailability}).
\begin{table}[htp]
\centering
\caption{Overview of photometric time series observations of TOI-2580}
\begin{tabular}{c|c|c|c|c}
Date & Tel. size & Band & Image scale & Apertures\\
\hline \hline
2021-10-17 &  0.3\,m & Rc & $1\farcs15$ & $10\farcs1$ \\
2022-03-12 & 0.3\,m & B & $0\farcs712$ & $3\farcs87$ \\
\end{tabular}
\end{table}
\\
We observed a full transit window of TOI-2580.01 in Rc band on UTC 17 October 2021 from the Astronomical Observatory University of Siena in Siena, Italy. The 0.3\,m telescope is equipped with a $3072\times2048$ SBIG STL-6303E detector, which  has an image scale of $1\farcs15$ per pixel, resulting in a $59\arcmin\times39\arcmin$ FoV. The differential photometric data were extracted using {\tt AstroImageJ} \citep{Collins:2017}. We used circular photometric apertures with a radius of $10\farcs1$. The target star aperture included all of the flux from the nearest known neighbor in the Gaia DR3 catalog (Gaia DR3 490410811948906368), which is $\sim4\farcs3$ northwest of TOI-2580.
\\
We observed a nearly full transit window of TOI-2580.01 in B filter on UTC 12 March 2022 from the Silesian University of Technology Observatory (SUTO) 0.3\,m telescope is located near Pyskowice, Poland. The telescope is equipped with a $4007\times2671$ pixel Atik 11000M detector, which has an image scale of 0$\farcs$712~pixel$^{-1}$, resulting in a $38\arcmin\times26\arcmin$ FoV. The differential photometric data were extracted using {\tt AstroImageJ}.  We used circular photometric apertures with a radius of $3\farcs87$, which excluded most of the flux from the nearest known neighbor in the Gaia DR3 catalog.
The light curve data are available on the {\tt EXOFOP-TESS} website\footnote{\href{https://exofop.ipac.caltech.edu/tess/target.php?id=102713734}{https://exofop.ipac.caltech.edu/tess/target.php?id=102713734}} and are included in the global modeling described in Sect. \ref{Sec:AnalysisJointFit}.

\subsubsection*{TOI-6016} \label{Sec:6016_GroundBasedPHot}


There are five (nearly) full transits observed of TOI-6016.  The light curves are available online (see Sect. \ref{Sec:DataAvailability}).
\begin{table}[htp]
\centering
\caption{Overview of photometric time series observations of TOI-6016}
\begin{tabular}{c|c|c|c|c}
Date & Tel. size & Band & Image scale & Apertures\\
\hline \hline
2023-08-10 &  0.3\,m & R & $1\farcs97$ & $13\farcs8$ \\
2023-08-18 & 0.3\,m & R & $0\farcs68$ & $5\farcs4$\\
2023-09-19 & 0.3\,m & g' & $0\farcs712$ & $7\farcs1$\\
2023-09-20 & 0.36\,m & g' & $1\farcs00$ & $7\farcs$\\
2023-10-14 & 0.36\,m & B, I & $0\farcs95$ & $6\farcs6$, $5\farcs7$ \\
\end{tabular}
\end{table}
\\
We observed a full transit window of TOI-6016.01 in R on UTC 10 August 2023 from the Thüringer Landessternwarte Tautenburg (TLS) Tautenburg Exoplanet Search Telescope (TEST) \citep{TEST2009} in Tautenburg, Germany. The telescope is equipped with an $4096\times4096$ a Moravian Instruments G4 16000 CCD detector with an image scale of 2$\farcs$~pixel$^{-1}$. The differential photometric data were extracted using {\tt AstroImageJ}. We used circular photometric apertures with a radius of $13\farcs8$, which excluded the flux from the nearest known neighbor in the Gaia DR3 catalog (Gaia DR3 428359976320449024), which is $\sim26\farcs4$ southwest of TOI-6016.
\\
We observed a full transit window of TOI-6016.01 in R on UTC 18 August 2023 from The Observatori Astronòmic de la Universitat de València (OAUV-TURIA2), located near Valencia, Spain. The OAUV-TURIA2 0.3\,m telescope is equipped with a QHY600 detector, which has an image scale of $0\farcs68$~pixel$^{-1}$, resulting in a $109\arcmin\times73\arcmin$ FoV. The differential photometric data were extracted using {\tt AstroImageJ}. We used circular photometric apertures with a radius of $5\farcs4$, so we excluded the flux from the nearest (distance $26.4\farcs$) known neighbor in the Gaia DR3 catalog.
\\
We observed a full transit window of TOI-6016.01 in Sloan $g'$ on UTC 19 September 2023 from Hamilton College Observatory in Clinton, NY, USA.  The 0.3\,m telescope is equipped with a ZWO ASI 1600 detector, which has an image scale of $0\farcs712$~pixel$^{-1}$, resulting in a $27.6\arcmin\times20.9\arcmin$ FoV. The differential photometric data were extracted using {\tt AstroImageJ}. We used circular photometric apertures with a radius of $7\farcs1$, which included the flux from the nearest known neighbor in the Gaia DR3 catalog.
\\
We observed a full transit window of TOI-6016.01 in Sloan $g'$ on UTC 20 September 2023 from the Acton Sky Portal private observatory in Acton, MA, USA. The 0.36\,m telescope is equipped with an SBIG ST-8300M camera having an image scale of $1\farcs00$ per pixel, resulting in a $17\times17\arcmin$ FoV. The image data were calibrated and photometric data were extracted using {\tt AstroImageJ}. We used a circular photometric aperture with a radius of $7\arcsec$ centered on TOI-6016, which excluded all of the flux from the nearest known neighbor in the Gaia DR3 catalog.
\\
We observed a nearly full transit window of TOI-6016.01 in alternating B and I bands on UTC 14 October 2023 from the Villa\,'39 observatory in Landers, California. The 0.36\,m telescope is equipped with a SX-56 camera with an image scale of $0\farcs95$ per $2\times2$ binned pixel, resulting in a $33\farcm1\times33\farcm1$ FoV. The image data were calibrated and photometric data were extracted using {\tt AstroImageJ}. We used a circular photometric aperture on TOI-6016 a radius
of $6\farcs6$ for B band and $5\farcs7$ for the I band, which excluded all of the flux from the nearest known neighbor in the Gaia DR3 catalog.
The light curve data are available on the {\tt EXOFOP-TESS} website\footnote{\href{https://exofop.ipac.caltech.edu/tess/target.php?id=327369524}{https://exofop.ipac.caltech.edu/tess/target.php?id=327369524}} and  included in the global modeling process described in Sect. \ref{Sec:AnalysisJointFit}. 

\subsubsection*{TOI-6130} 


There were three full transits observed for TOI-6130. The light curves are available online (see Sect. \ref{Sec:DataAvailability}).
\begin{table}[htp]
\centering
\caption{Overview of photometric time series observations of TOI-6130}
\begin{tabular}{c|c|c|c|c}
Date & Tel. size & Band & Image scale & Apertures\\
\hline \hline
2023-06-17 & 0.51\,m & Rc & $1\farcs77$ & $5\farcs8$ \\
2023-07-04 & 0.35\,m & g' & $0\farcs73$ & $6\farcs6$ \\
2023-08-15 & 1.52\,m & g, r, i, $z_s$ & $0\farcs44$ & $4\farcs4$ \\
\end{tabular}
\end{table}
\\
We observed a full transit window of TOI-6130.01 in Rc on UTC 17 June 2023 from the Evans 0.51\,m telescope located at the El Sauce Observatory in Coquimbo Province, Chile. The telescope is equipped with a $1536\times1024$ pixel SBIG STT-1603-3 detector, with an image scale of $1\farcs77$ per $2\times2$ binned pixel, resulting in an $18.8\arcmin\times12.5\arcmin$ FoV. The differential photometric data were extracted using {\tt AstroImageJ}. We used circular photometric apertures a radius
of $5\farcs8$, which excluded the flux from the nearest known neighbor in the Gaia DR3 catalog  (Gaia DR3 2736716258652620544), which is $\sim10\farcs3$ south of TOI-6130.
\\
We observed a full transit window of TOI-6130.01 in Sloan $g'$ on UTC 4 July 2023 from the Las Cumbres Observatory Global Telescope (LCOGT) \citep{Brown:2013} network node at South Africa Astronomical Observatory near Cape Town, South Africa (SAAO). The 0.35\,m Planewave Delta Rho 350 telescope is equipped with a $9576\times6388$ QHY600 CMOS camera, with an image scale of $0\farcs73$ per pixel, resulting in a $114\arcmin\times72\arcmin$ FoV. However, the image data were collected using the $30\arcmin\times30\arcmin$ central FoV sub-frame mode. The images were calibrated by the standard LCOGT {\tt BANZAI} pipeline \citep{McCully:2018} and differential photometric data were extracted using {\tt AstroImageJ}. We used circular photometric apertures a radius
of $6\farcs6$, which excluded the flux from the nearest known neighbor in the Gaia DR3 catalog.
\\
The most recent transit was observed on 15 August 2023 using the multiband imager MuSCAT2 \citep{Narita2019},  mounted on the 1.52 m TCS telescope at the Teide Observatory in the Canary Islands, Spain. MuSCAT2 has four optical channels each equipped with a 1k$\times$1k CCD camera with a pixel scale of 0."44~pixel$^{-1}$, providing a FoV of 7.'4 × 7.'4; it is is capable of simultaneous imaging in g, r, i, and $z_s$ bands. A dedicated MuSCAT2 pipeline \citep{MUSCAT2transitpipeline} was used to perform the data reduction and to extract the photometry. The optimal light curves were obtained using the three brightest comparison stars and an uncontaminated 4.4" aperture radius. The transit was detected on-time and on-target with $(R_p/R_\ast)^2$ transit depths of 12.8, 12.5, 12.3, 12.0 ppt for g, r, i, and $z_s$ bands, respectively.
The light curve data are available on the {\tt EXOFOP-TESS} website\footnote{\href{https://exofop.ipac.caltech.edu/tess/target.php?id=210083929}{https://exofop.ipac.caltech.edu/tess/target.php?id=210083929}} and are also included in the global modeling described in Sect. \ref{Sec:AnalysisJointFit}.

\subsection{High angular-resolution observations} \label{Sec:HiAngResObs}

For all four targets, there were several ground-based observations of the targets with high angular-resolution, both using adaptive optics systems and speckle imaging.
These can be used to identify nearby sources that may contaminate the TESS photometry, resulting in an underestimated planetary radius. Alternatively, they may be the source of astrophysical false positives, such as background eclipsing binaries.

\subsubsection*{TOI-1295} \label{Sec:1295HighRes}

\begin{figure}[htp]
    \centering
    \includegraphics[width=\linewidth]{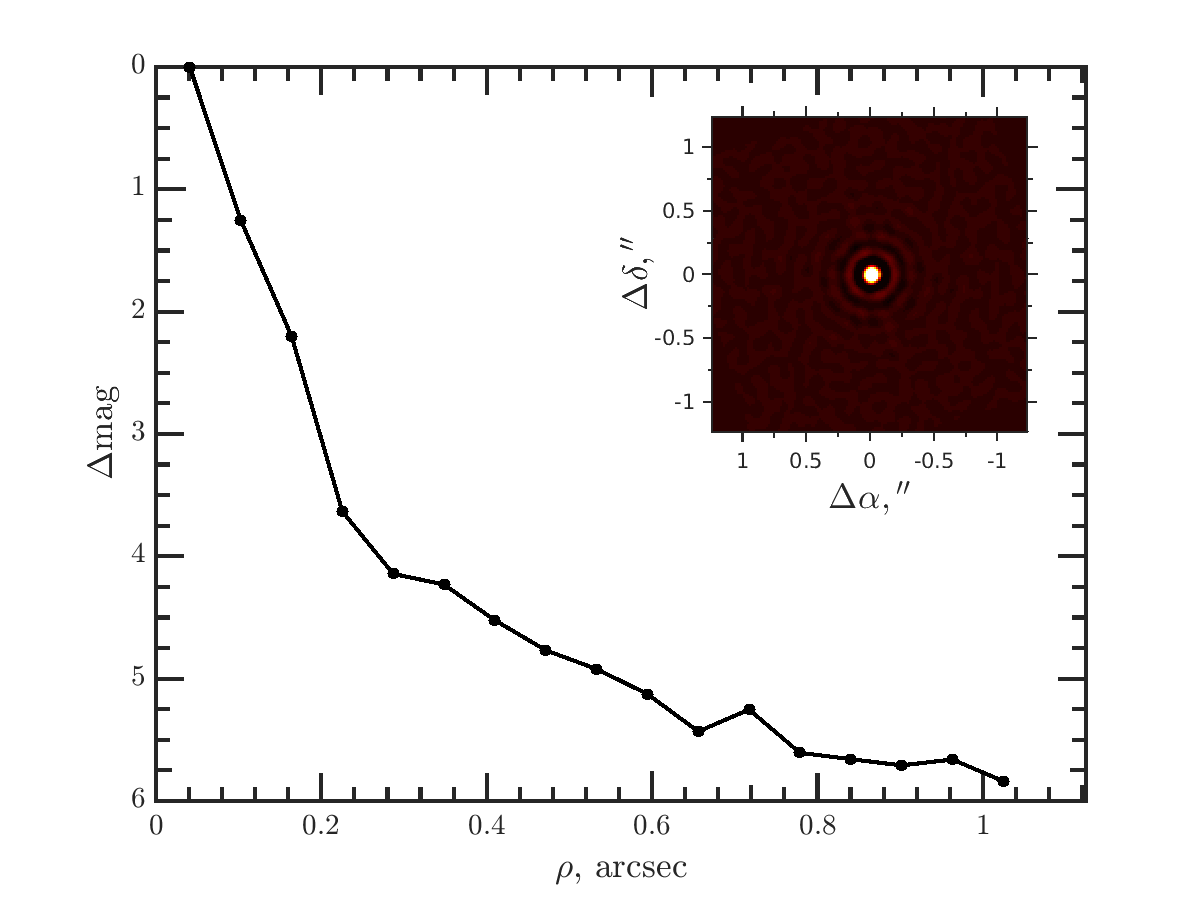}
    \caption{Speckle sensitivity curve and auto correlation function (ACF) of TOI-1295 obtained with the SAI Speckle polarimeter.}
    \label{fig:1295_Speckle}
\end{figure}

TOI-1295 was observed on 2 February 2021 with the Speckle Polarimeter \citep{2017Safonov} on the 2.5~m telescope at the Caucasian Observatory of Sternberg Astronomical Institute (SAI) of Lomonosov Moscow State University (see Fig. \ref{fig:1295_Speckle}). Electron Multiplying CCD Andor iXon 897 was used as a detector. The atmospheric dispersion compensator allowed observation through the wide-band $I_c$ filter. The power spectrum was estimated from 4000 frames with 30~ms exposure. The detector has a pixel scale of $20.6$~mas~pixel$^{-1}$, and the angular resolution was 89~mas. Long-exposure seeing was $1\farcs45$. No stellar companions brighter than $\Delta I_C=4.2$~mag and $5.8$~mag at $\rho=0\farcs25$ and $1\farcs0$, respectively, were detected (where $\rho$ is the separation between the source and the potential companion).
\\
Additionally, two imaging observations were obtained using NIR adaptive optics. 
They both were acquired with the ShARCS camera on the Shane 3-m telescope at Lick Observatory on Mt. Hamilton, USA \citep{2012SPIE.8447E..3GK, 2014SPIE.9148E..05G, 2014SPIE.9148E..3AM} in $K_s$-band ($\lambda_0 = 2.150$~$\mu$m, $\Delta \lambda = 0.320$~$\mu$m), and $J$-band ($\lambda_0 = 1.238$~$\mu$m, $\Delta \lambda = 0.271$~$\mu$m).
No stellar companions with $\Delta J \le 4$~mag and $\Delta K_s \le 4$~mag within $1\farcs0$ were found (Dressing et al. in prep).

\subsubsection*{TOI-2580}

For TOI-2580 there are two imaging observations using NIR adaptive optics from ShARCS (see information above for TOI-1295). No stellar companions beyond those listed in the TIC were detected by the observations (Dressing et al., in prep).

\subsubsection*{TOI-6016}

\begin{figure}[htp]
    \centering
    \includegraphics[width=\linewidth]{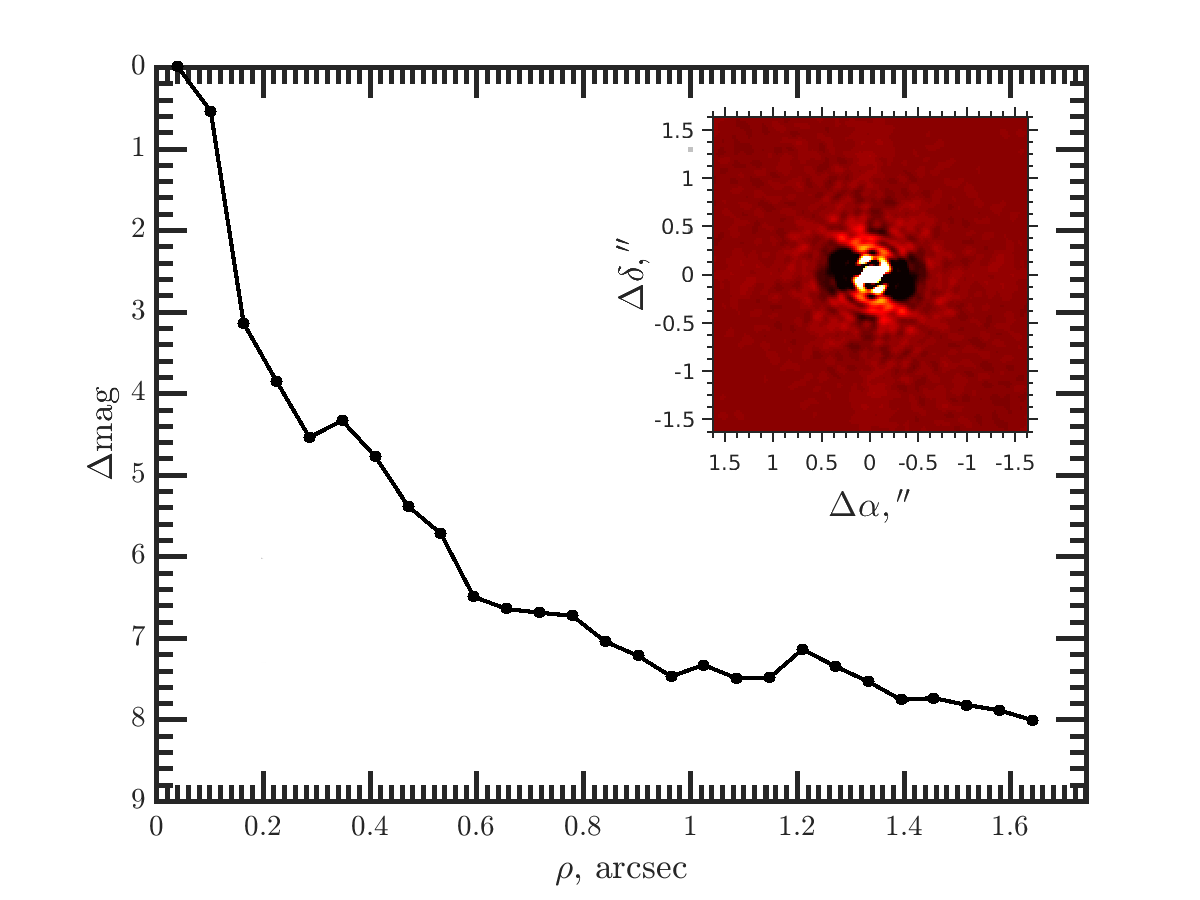}
    \caption{Speckle-sensitivity curve and ACF of TOI-6016 obtained with the 2.5 m telescope at the Caucasian Observatory of Sternberg (SAI).}
    \label{fig:6016_Speckle}
\end{figure}

TOI-6016 was observed on 24 December 2022 with the speckle polarimeter on the 2.5-m telescope at the Caucasian Observatory of Sternberg Astronomical Institute (SAI) of Lomonosov Moscow State University (see Fig. \ref{fig:6016_Speckle}). This speckle polarimeter uses the high-speed low-noise CMOS detector Hamamatsu ORCA-quest \citep{2017Safonov}. The atmospheric dispersion compensator was active, which allowed for the $I_\mathrm{c}$ band to be used. The respective angular resolution is 0.083", while the long-exposure atmospheric seeing was 0.54". There were no stellar companions detected that are brighter than $\Delta I_\mathrm{c}=4.5$ and 7.2 at $\rho$ = 0.25" and 1.0", respectively, where $\rho$ is the separation between the source and the potential companion.

\subsubsection*{TOI-6130}

\begin{figure}[htp]
    \centering
    \includegraphics[width=\linewidth]{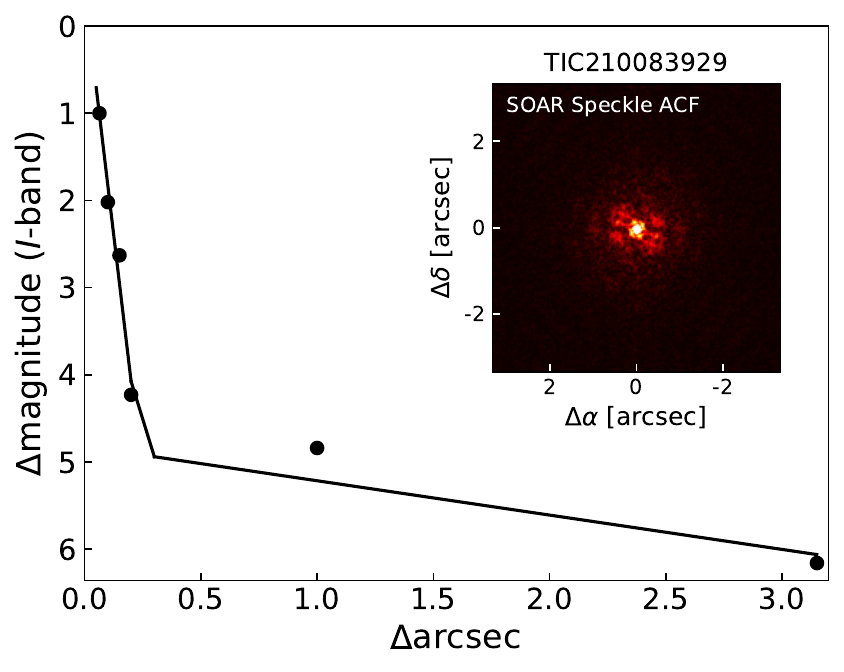}
    \caption{Speckle-sensitivity curve and ACF of TOI-6130 obtained with the SOAR 4.1 m telescope.}
    \label{fig:6130_Speckle}
\end{figure}

On 31 August 2023 speckle imaging on the 4.1-m Southern Astrophysical Research (SOAR) telescope \citep{Tokovinin2018} was conducted in the Cousins I-band (a similar visible bandpass as TESS) to search for stellar companions around TOI-6130. This observation was sensitive to a 4.9-magnitude fainter star at an angular distance of 1 arcsec from the target \citep[for more details of the observations within the SOAR TESS survey see][]{Ziegler2020}. The 5 $\sigma$ detection sensitivity and speckle auto-correlation functions from the observations are shown in Fig. \ref{fig:6130_Speckle}. No nearby stars were detected within 3$\arcsec$ of TOI-6130 in the SOAR observations.

\section{Spectroscopic observations} \label{Sec:SpecObs}

\subsection{Recon spectra} \label{Sec:SingleSpecObs}

For all the objects, we took several single spectroscopic observations. There were done mainly with the high-resolution echelle spectrograph HIRES (spectral resolution R $\sim$ 50000) at the 10 m Keck telescope on top of the Mauna Kea in Hawaii, USA, and the Tillinghast Reflector Echelle Spectrograph (TRES) (spectral resolution R $\sim$ 44000) at the 1.5~m Smithsonian Astrophysical Observatory's Fred L. Whipple Observatory on Mt. Hopkins in Arizona, USA. These data were used to derive the properties of the target stars and assess whether the target is generally suitable for precise radial velocity follow-ups. This step also made it possible to rule out some false positives stemming from eclipsing binaries.

\subsection{High-resolution time-series spectroscopy} \label{Sec:HiResTimeSeriesSpecObs}

To create a radial velocity curve and infer the planetary mass, high-resolution spectroscopic observations were taken for the four objects. All of them were observed with the Manfred-Hirt-Planet Spectrograph (MaHPS). For TOI-2580 and TOI-6016, there were additional observations with the astronomical spectrograph NEID. For TOI-6016, spectroscopic observations from TRES were used as well.

\subsubsection{Manfred-Hirt-Planet Spectrograph}

MaHPS \citep[described in][]{HannaDoktorarbeit, Pfeiffer1998} is one of the three instruments at the 2.1 m Fraunhofer telescope at the Wendelstein Observatory operated by the Ludwig-Maximilians-University in Munich. It consists of the Fiber Optics Cassegrain Échelle Spectrograph (FOCES) and a Laser Frequency Comb for precise wavelength calibration. Additionally, the spectra can be wavelength calibrated using a Thorium-Argon lamp for nights when the comb is not in use. During scientific operation, MaHPS can reach a radial velocity precision of a few m/s for bright stars.
\\
These four planets and those presented in Thomas et al (in prep.) are the first planetary candidates confirmed via mass measurements from RV data obtained with MaHPS. As the four targets are at the edge of the magnitude limit for MaHPS, we have selected an exclusion criterion for observations unsuited for precise RV work based on the S/N of the observations (see  the paragraph below on "Adaptions in the data analysis for faint targets"). We also decided to exclude ThAr-calibrated observations on days where observations were taken both with and without simultaneous comb-light because the latter has higher precision.\\

\subsubsection*{General data reduction}

To extract radial velocity values from the spectroscopic observations taken with MaHPS, two software packages were used. General AstronoMical Spectra Extractor \citep[{\tt GAMSE},][]{Wang2016} performs 2D to 1D spectra extraction and a first ThAr wavelength calibration. Munich Analyzer for Radial velocity Measurements with b-spline Optimized Templates \citep[{\tt MARMOT},][]{Kellermann2020} proceeds with a Laser frequency comb wavelength calibration and an RV extraction.
In the following, we briefly describe the analysis steps that were performed in general on the target spectra to obtain radial velocities (for details see Thomas et al (in prep)).
\\
As a first step, the analysis software {\tt GAMSE} calculates  the 1D spectra from the raw
2D spectra for each order, together with a first wavelength calibration. In this process, first, an image is created by {\tt GAMSE} that has been treated with the general data reduction steps that are common for every astronomical observation. Those include the subtraction of the overscan and the bias, flat correction using a sensitivity map, and correcting the background for stray light. From this corrected 2D spectrum, {\tt GAMSE} extracts a 1D spectrum. The flux along one order is summed up while weighting the flux inside each pixel in case they are located at the edges of the order.
Lastly, {\tt GAMSE} performs a first ThAr wavelength calibration and identifies the number of the orders. In this step, the lines found in the observed ThAr spectra are compared with information about emission lines of ThAr from the literature. Thus, each line can be assigned to a wavelength, and this wavelength is allocated to a certain pixel.
\\
\\
Next, the program {\tt MARMOT} is used to get radial velocity values for the taken spectra of the objects. First, artificial spikes (caused by cosmics or Hot pixels) from the spectra are removed.
Second, the wavelength calibration is improved and refined by using comb light that is observed simultaneously. It has experienced the same environmental variations, and, thus, the same changes in the observed spectra, so one accounts for shifts in the spectrum induced by instrumental effects.
This is followed by the barycenter correction.
\\
Next, a template spectrum is created that is used as a reference for each observed spectrum when calculating the radial velocity values. Here, several observed spectra (with high S/N) are combined with a B-spline optimization algorithm, which constructs a template spectrum that is less affected by artificial structure in the spectra, such as noise or cosmic rays.
\\
It is also important to exclude certain pre-defined regions from the spectrum. These regions are mainly determined by specifications of the Wendelstein observatory site, instruments, or the Earth’s atmosphere. They are independent of the observed star. Examples are regions containing broad absorption lines, for example the H$\alpha$-line, known telluric lines, and the detector edges.
\\
Lastly, the radial velocities for the separate frames are computed. The template spectrum is fitted using a $\chi^2$-fit to the order of the science spectrum, which gives a value of the wavelength shift for each order. By weighting the wavelength shift of each order with the error from the fit, the final radial velocity values are calculated.\\

\subsubsection*{Adaptions in the data analysis for faint targets}

The objects presented here were the first targets of V-band magnitudes higher than 11 that were spectroscopically observed with MaHPS to obtain mass measurements. 
Thus, we developed certain criteria to ensure efficient observations for the future and exclusion criteria that mark low-quality data (e.g., due to bad weather conditions or technical issues). As the signal-to-noise ratio (S/N) of a given observation can be a good indicator of the observing conditions, we decided to only include spectroscopic data above a certain S/N limit. A reduced number of photons resulting in a lower S/N is normally caused by either bad weather, for example clouds, a low inclination, or problems with the guiding.
We used the order 86 (659.5~nm - 669.0~nm) as the reference instead of the mean-weighted S/N of the total frame. There are mainly two motivations for this. First, it is already available for each frame after the basic reduction steps performed by {\tt GAMSE} and before specific functions of the program {\tt MARMOT} are run; for instance, even before the radial velocity values are computed for each frame.
Second, the two S/N limit definitions (S/N of order 86 and the mean-weighted S/N) have been found to lie on approximately linear relation (close to 1:1, see a figure shown online (Link in Sect. \ref{Sec:DataAvailability})). Then, the distribution of S/N through the various orders is similar for all frames and the S/N (in order 86) normally shows one of the largest S/N values for the whole set of values.
\\ 
A good balance between maximizing the available RV data, but also limiting the data to high-quality ones, was found by excluding the "low tail" of the S/N values. As visualized in a figure shown online (see Sect. \ref{Sec:DataAvailability}), we have excluded all frames with an S/N in order 86 below 14 (for TOI-1295), 12 (for TOI-2580 and TOI-6016), and 13 (for TOI-6130). The thresholds here are different as the targets also have different magnitudes with TOI-1295 being the brightest. 
For TOI-1295 specifically, the number of excluded observations is the highest for two main reasons: during the scope of its observation the observation strategy was adapted for fainter targets; thus, several observations also served as a learning procedure. Additionally, there were several attempts to measure the Rossiter-McLaughlin (RM) effect. For this purpose, there were observations taken during transits even if the observing conditions were unfavorable, for example low altitude of the target or cloudy sky conditions. Thus, for TOI-1295 from the initial number of observations, about 26\% were excluded; whereas for TOI-2580, TOI-6016, and TOI-6130, it was less than 20 $\%$.
In a few cases (five frames of TOI-1295, four frames of TOI-2580, none for TOI-6016, and five for TOI-6130), we decided to exclude ThAr-calibrated observations on days where observations were taken both with and without simultaneous comb-light, as the latter has higher precision. 

\subsubsection{NEID}

The NASA/NSF Extreme Precision Doppler Spectrometer instrument concept NEID (NN-EXPLORE Exoplanet Investigations with Doppler spectroscopy) is a high-resolution echelle spectrograph at the 3.5~m WIYN telescope at Kitt Peak National Observatory in Arizona, USA. 
It is operating in the wavelength range of 380-930~nm and its single-measurement radial velocity precision is $\sim$ 27~cm/s \citep{NEID_Halverson2016}.
The observations were taken in high-resolution (HR) mode, with spectral resolution $R \sim 110{,}000$.
The NEID spectra were reduced and precise RVs extracted through cross-correlation with a stellar line mask \citep{Baranne1996,Pepe2002} appropriate to the targets' stellar type, as implemented in v1.3.0 of the standard NEID Data Reduction Pipeline (DRP).\footnote{\url{https://neid.ipac.caltech.edu/docs/NEID-DRP/}}

\subsubsection{TRES}

TRES \citep[][]{Furesz,TRES2012} is an optical fiber-fed echelle spectrograph located at the Fred Lawrence Whipple Observatory (FLWO) at Mt. Hopkins, Arizona. It has a resolving power of R $\sim$ 44,000 and operates in the wavelength range 390-910~nm.
All spectra taken with TRES were visually inspected to check for composite spectra and then a multi-order spectral analysis was performed following procedures outlined in \cite{Buchhave2010} and \cite{Quinn2012}. Essentially, the spectra were cross-correlated order-by-order against a median combined template to derive radial velocities. The spectra were then corrected for zero-point offsets using a history of standard star observations.

\subsubsection{Spectroscopic observations of the target stars}

\begin{table}[htp]
\centering
\caption{Overview of the spectroscopic observations}
\begin{tabular}{c|c|c}
 & Spectrograph & Number of RVs\\
\hline \hline
TOI-1295 & MaHPS & 135\\
\hline
TOI-2580 & MaHPS & 31\\
& NEID & 6\\
\hline
TOI-6016 & MaHPS & 26\\
& NEID & 6\\
& TRES & 16\\
\hline
TOI-6130 & MaHPS & 42\\
\end{tabular}
\end{table}

\subsubsection*{TOI-1295}

TOI-1295 was observed exclusively with MaHPS. As this was the very first planetary candidate to be confirmed with the MaHPS, a larger number of observations are available. In total, we used 135 (108 without transit frames) observations on 42 nights (1 May 2022 - 2 February 2023), with exposure times of 1800~s with a minimum S/N of 14 in the order 86.
\\
TOI-1295 was deliberately chosen as a target to not only measure the mass of the planet candidate but also incorporate spectroscopic observations during transit times and thus use the Rossiter-McLaughlin effect to obtain the projected obliquity of the star (work in progress). However, we have not incorporated these observations and their analysis here due to the low number and quality of the transit frames.
With TOI-1295 we aimed to move toward observing fainter stars (in the V-band a magnitude $\geq$ 11) that have been observed with the MaHPS before. Using the RV method, we found that for TOI-1295 with a V-band magnitude of 11.3~mag a mass precision of better than 15$\%$ can be achieved (planetary mass: 1.5 $\pm$ 0.2~$\text{M}_{Jup}$).

\subsubsection*{TOI-2580}

To create the radial velocity curve for TOI-2580, spectroscopic data from the MaHPS and NEID were used.
From MaHPS, there were 31 observations on 12 nights (14 February 2023 to 25 September 2023), with exposure times of 1800~s with a minimum S/N of 12 in order 86 used.
Additionally, there were six observations (22 November 2022 to 23 April 2023), with exposure times ranging from 300~s to 420~s taken with the NEID spectrograph. 

\subsubsection*{TOI-6016}

For creating the radial velocity curve for TOI-6016 spectroscopic data from the MaHPS, TRES, and NEID were used.
From MaHPS, there were 26 observations on 10 nights (12 August 2023 - 3 December 2023), with exposure times of 1800~s, with a minimum S/N of 12 in order 86 were used.
Additionally, there are 16 velocities obtained with data from the TRES spectrograph.
Two initial reconnaissance spectra were observed on 25 December 2022 and 13 January 2023; and then additional fourteen spectra were acquired between 7 September 2023 and 10 October 2023, with typical exposure times of 1000~s.
Furthermore, there were six observations (3 October 2023 - 21 October 2023) with exposure times of 480~s taken with the NEID spectrograph. 

\subsubsection*{TOI-6130}

To create the radial velocity curve for TOI-6130, only data from the MaHPS were used. There were 42 observations on 19 nights (10 July 2023 - 20 September 2023), with exposure times of 1800~s with a minimum S/N of 13 in order 86 used. The tables containing all RV values used in this analysis are available in electronic form at the CDS \footnote{anonymous ftp to \href{cdsarc.u-strasbg.fr}{cdsarc.u-strasbg.fr} (130.79.128.5), \\ or \href{http://cdsweb.u-strasbg.fr/cgi-bin/qcat?J/A+A/}{http://cdsweb.u-strasbg.fr/cgi-bin/qcat?J/A+A/}}.

\section{Stellar characterization}\label{Sec:StellarCaracterization}

\subsection{Stellar parameters from spectral analysis} \label{Sec:StellarCharacterizationSpectralAnalysis}

We first derived  the spectroscopic stellar parameters using the MaHPS and TRES observations. The resulting values were then used as input values for SED-fitting for the bulk stellar parameters.\\


\subsubsection{Analysis of the TRES data with the {\tt SPC} tool}

For TRES spectra, the stellar effective temperature, surface gravity, metallicity, and rotational velocity were derived by cross-correlating the spectra against Kurucz atmospheric models \citep{Kurucz1992} and using the {\tt Stellar Parameter Classification} tool; \citep[{\tt SPC}][]{Buchhave2014, Buchhave2012}.\\

\subsubsection{Analysis of MaHPS data with {\tt SpecMatch-Emp}}
To get the stellar mass and radius, effective temperature, surface gravity, metallicity, and an estimate of the age, we performed an analysis of the spectra from MaHPS with the Python package {\tt SpecMatch-Emp} \citep{SpecMatch2017}.
The keys steps are summarized in the following. We first normalized the template spectrum of the observed star and shifted it to match the wavelength region of the dense spectral library of high resolution (R $\sim$ 55,000), high S/N ($>$ 100) spectra taken with Keck/HIRES by the California Planet Search \citep{SpecMatch2017}.
From this library, five reference stars are picked whose properties match the target star closely using a cross-correlation of the spectra. In the next step, the best-fitting linear combination of the properties of the reference stars is selected to determine the target star's parameters. For more information, we refer to  \citet{SpecMatch2017}.\\
The resulting values from MaHPS are compared with the average (+ maximum error) of the stellar parameters derived from the TRES data (see Tables \ref{tab*:ComparisonStellarParameters1} and \ref{tab*:ComparisonStellarParameters2}). They generally agree within the error bars, except for the metallicity of TOI-2580. For the SED fitting, we decided to use the weighted averaged values (see Table \ref{tab:StellarParameters}) obtained from these two values as input values for the fitting process and for the conversion of the obtained values into physical units. Additionally, we gave a first estimate for the stellar radius obtained from the MaHPS spectra and a value for $v\sin(i)$ from the TRES spectra.

\subsubsection{Analysis of the NEID data with {\tt SpecMatch-Emp}}

The NEID data (for TOI-2580 and TOI-6016) were analyzed similarly as the MaHPS with {\tt SpecMatch-Emp} \citep{SpecMatch2017}. In both cases, we shifted each observation onto the barycentric rest frame and stacked them into a single template spectrum before running the code.

\begin{table*}[htp]
    \centering
    \caption{Comparison of derived stellar parameters from the MaHPS, TRES, and NEID spectra for TOI-1295 and TOI-2580}
    \begin{tabular}{l||c|c||c|c|c}
    & \multicolumn{2}{c||}{TOI-1295} & \multicolumn{3}{c}{TOI-2580}\\
    \hline \hline
    & MaHPS & TRES & MaHPS & TRES & NEID \\
    \hline
    $T_{\text{eff}}$ [K] & 6269 $\pm$ 110 & 6280 $\pm$ 60 & 6091 $\pm$ 110 & 6140 $\pm$ 80 & 6080 $\pm$ 110\\
    log g & 4.11 $\pm$0.12 & 4.1 $\pm$ 0.1 & 4.11 $\pm$ 0.12 & 4.20 $\pm$ 0.14 & - \\
    $[$Fe/H$]$ & 0.16 $\pm$ 0.09 & 0.26 $\pm$ 0.08 & 0.08 $\pm$ 0.09 & 0.41 $\pm$ 0.08 & 0.21 $\pm$ 0.09\\
    \hline
    R$_*$ [$R_{\odot}$] & 1.74 $\pm$ 0.18 & - & 1.64 $\pm$ 0.18 & - & 1.5 $\pm$ 0.3 \\
    $v\sin(i)$ [km/s] & - & 7.8 $\pm$ 0.5 & - & 7.4 $\pm$ 0.5 & - \\
    \end{tabular}    \label{tab*:ComparisonStellarParameters1}
\end{table*}

\begin{table*}[htp]
    \centering
    \caption{Comparison of derived stellar parameters from the MaHPS, TRES, and NEID spectra for TOI-6016 and TOI-6130}
    \begin{tabular}{l||c|c|c||c|c}
    & \multicolumn{3}{c||}{TOI-6016} & \multicolumn{2}{c}{TOI-6130}\\
    \hline \hline
    & MaHPS & TRES & NEID & MaHPS & TRES\\
\hline
    $T_{\text{eff}}$ [K] & 6004 $\pm$ 110 & 6130 $\pm$ 50 & 5970 $\pm$ 110 & 5816 $\pm$ 110 & 5960 $\pm$ 50\\
    log g & 4.10 $\pm$ 0.12 & 4.26 $\pm$ 0.10 & - & 4.19 $\pm$ 0.12 & 4.37 $\pm$ 0.10 \\
    $[$Fe/H$]$ & 0.21 $\pm$ 0.09 & 0.33 $\pm$ 0.08 & 0.28 $\pm$ 0.09 & 0.17 $\pm$ 0.09 & 0.19 $\pm$ 0.08\\
    \hline
    R$_*$ [$R_{\odot}$] & 1.70 $\pm$ 0.18 & - & 1.9 $\pm$ 0.4  & 1.27 $\pm$ 0.18 & - \\
    $v\sin(i)$ [km/s] & - & 8.4 $\pm$ 0.5 & & - & 5.3 $\pm$ 0.5\\
    \end{tabular}    \label{tab*:ComparisonStellarParameters2}
\end{table*}

\subsection{Stellar parameters from SED-fitting} \label{Sec:StellarCharacterizationSEDFit}

\begin{figure*}[htp]
    \centering
    \includegraphics[width=0.49\textwidth]{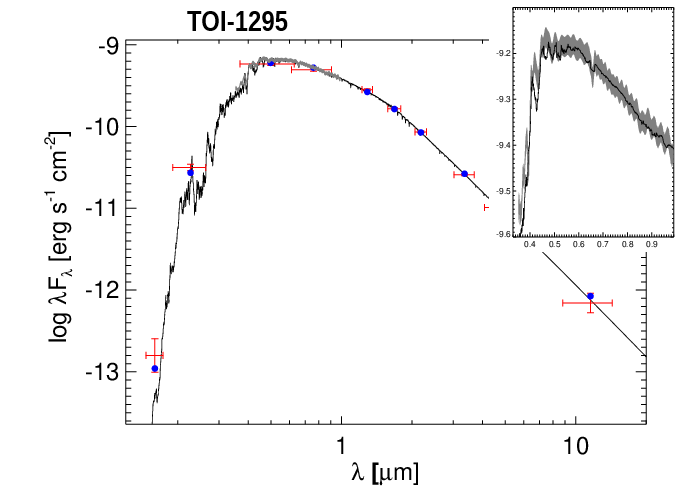}
    \includegraphics[width=0.49\textwidth]{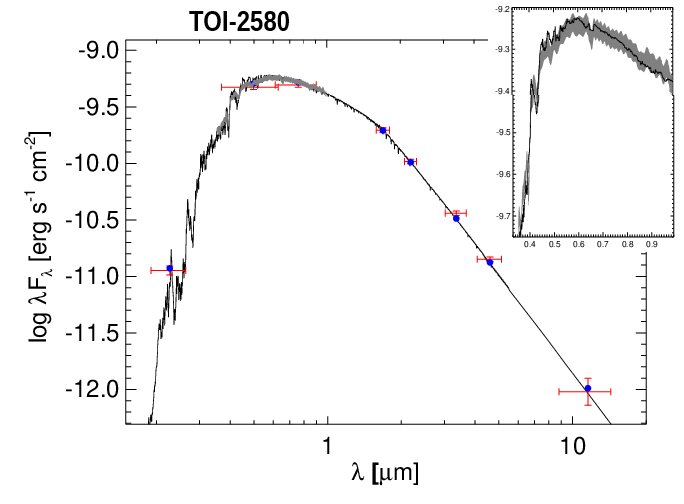}
    \includegraphics[width=0.49\textwidth]{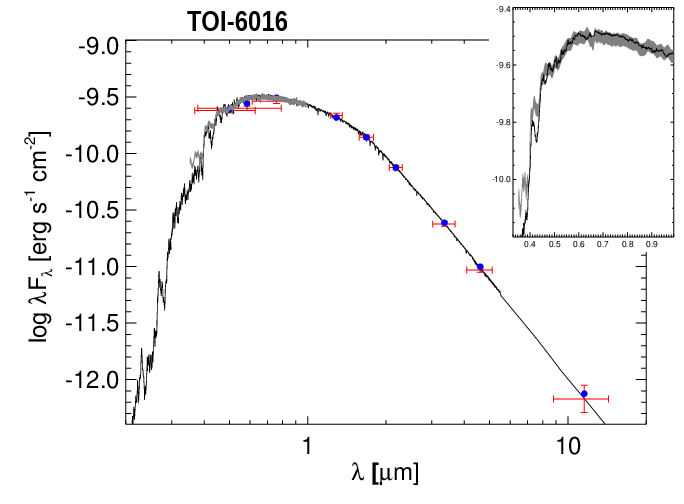}
    \includegraphics[width=0.49\textwidth]{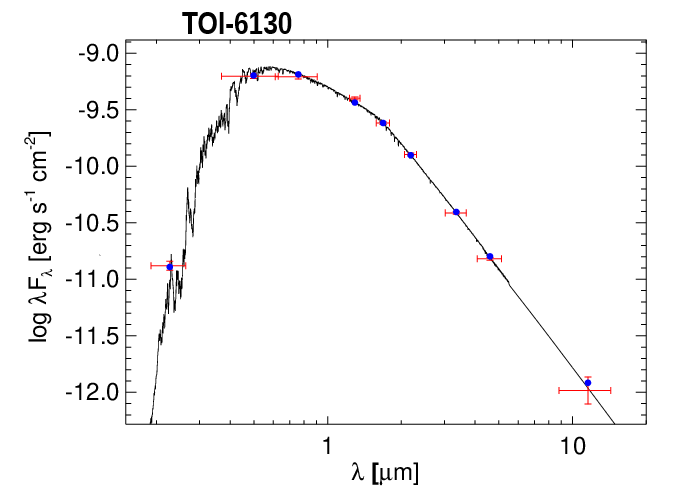}
    \caption{Spectral energy distributions of the four host stars. Top left: TOI-1295, top right: TOI-2580. Bottom left: TOI-6016. Bottom right: TOI-6130. Red symbols represent the observed photometric measurements, where the horizontal bars represent the effective width of the passband. Blue symbols are the model fluxes from the best-fit PHOENIX atmosphere model (black). The absolute flux-calibrated {\it Gaia\/} spectrum is shown as a gray swathe in the inset Fig..} \label{fig:sed}
\end{figure*}

For each star, we performed an analysis of the broadband spectral energy distribution (SED) of the star together with the {\it Gaia\/} DR3 parallax, in order to determine an empirical measurement of the stellar radius \citep{Stassun:2016,Stassun:2017,Stassun:2018}. Where available, we pulled the $JHK_S$ magnitudes from {\it 2MASS},  W1--W4 magnitudes from {\it WISE},  FUV and NUV magnitudes from {\it GALEX}, and $G_{\rm BP} G_{\rm RP}$ magnitudes from {\it Gaia}. We also utilized the absolute flux-calibrated {\it Gaia\/} spectrum where available. Together, the available photometry spans the full stellar SED over the wavelength range of 0.4--10~$\mu$m (TOI-6016) and 0.2--22~$\mu$m for the other three targets (see Fig.~\ref{fig:sed}).
\\
\\
We performed a fit using PHOENIX stellar atmosphere models \citep{Husser:2013}, adopting from the spectroscopic analysis the effective temperature ($T_{\rm eff}$), metallicity ([Fe/H]), and surface gravity ($\log g$). We fitted for the extinction ($A_V$), limited to the maximum line-of-sight value from the Galactic dust maps of \citet{Schlegel:1998}. Integrating the (unreddened) model SED gives the bolometric flux at Earth ($F_{\rm bol}$). Taking the $F_{\rm bol}$ together with the {\it Gaia\/} parallax and $T_{\rm eff}$ gives the stellar radius ($R_\star$). In addition, we estimated the stellar mass using the empirical relations of \citet{Torres:2010}.
We used the values for $v\sin(i)$ from ExoFOP together with the obtained stellar radius to get an estimate for P$_\text{rot}$/$\sin(i)$.
\\
Finally, where possible, we estimated the chromospheric activity index ($R'_{\rm HK}$) from the {\it GALEX} FUV and/or NUV excess via the empirical relations of \citet{Findeisen:2011}. We then estimated the stellar age via the empirical age-activity relations of \citet{Mamajek:2008}.
The resulting values for these parameters are shown in Table \ref{tab:StellarParameters}.
\\
We note for TOI-6130, the chromospheric activity predicts a rotation period that is $\sim$ 2x longer than that inferred from the combination of $v\sin(i)$, $R_\star$, and the assumption that $\sin(i) = 1$. There are known cases where the activity produces an increased line broadening and, therefore, a $v\sin(i)$ measurement that is larger than the true value. So one possibility is that the star is rotating more slowly in reality, which would then align with the R'hk activity-rotation-based estimate. In that case, it is a much older star ($\sim$ 10 Gyr) than implied by the relatively fast $v\sin(i)$ ($\sim$ 1 Gyr). The exact alternative values for TOI-6130 here are:
log Rhk (NUV) = -5.4 $\pm$ 0.1,
Rhk predicted P$_\text{rot}$ = 25.4 $\pm$ 2.2 d, and age (Rhk) = 10.1 $\pm$ 1.0 Gyr.

\begin{table*}[htp]
    \centering
    \caption{Stellar parameters of the four targets}
    \resizebox{\textwidth}{!}{
    \begin{tabular}{l c c c c c}
    Parameter & TOI-1295 & TOI-2580 & TOI-6016 & TOI-6130 & Ref.\\
    \hline \hline
    Identifiers\\
    ID (TOI) & 1295 & 2580 & 6016 & 6130\\
    ID (TYC) & 4421-02308-1 & 4076-01264-1 & 3665-00505-1 & 1696-01106-1\\
    ID (TIC) & 219852584 & 102713734 & 327369524 & 210083929\\
    \hline \hline
    Astrometric properties\\
    RA [hh:mm:ss.ss] & 17:06:41.348 & 04:08:47.782 & 00:18:05.766 & 22:36:49.200 & (1)\\
    Dec [dd:mm:ss.ss] & +67:52:17.278 & +67:06:59.573 & +59 45 58.342 & +16:49:37.753 & (1)\\
    Epoch (ICRS) & J2000 & J2000 & J2000 & J2000 & (1) \\
    Proper motion in RA [mas/yr] & -7.199 $\pm$ 0.020 & 10.607 $\pm$ 0.014 & -7.728 $\pm$ 0.010 & 18.435 $\pm$ 0.014 & (1)\\
    Proper motion in Dec [mas/yr] & 8.555 $\pm$ 0.021 & -5.049 $\pm$ 0.021 & -1.837 $\pm$ 0.011 & 9.357 $\pm$ 0.015 & (1)\\
    Parallax [mas] & 2.4656 [0.015] A & 2.6398 [0.0230] A & 2.7492 [0.0109] A & 4.5493 [0.0137] A & (1)\\
    Distance [pc] & 388 $\pm$ 5 & 379 $\pm$ 5 & 365 $\pm$ 4 & 221 $\pm$ 3 & ExoFOP\\
    \hline \hline
    Photometric properties \\
    B [mag] & 11.61 $\pm$ 0.15 & 12.11 $\pm$ 0.17 & 12.9 $\pm$ 0.4 & 11.76 $\pm$ 0.16 & (2)\\
    V [mag] & 11.303 $\pm$ 0.011 & 11.419 $\pm$ 0.012 & 11.886 $\pm$ 0.029 & 11.21 $\pm$ 0.011 & (2)\\
    TESS T [mag] & 10.797 $\pm$ 0.007 & 10.806 $\pm$ 0.006 & 11.386 $\pm$ 0.006 & 10.558 $\pm$ 0.006 & (2)\\
    Gaia G [mag] &  11.1295 $\pm$ 0.0028 &  11.2568 $\pm$ 0.0028 & 11.9100 $\pm$ 0.0028 & 10.9769 $\pm$ 0.0028 & (1)\\
    J [mag] & 10.341 $\pm$ 0.019 & - & 10.629 $\pm$ 0.021 & 9.966 $\pm$ 0.021 & (3)\\
    H [mag] & 10.123 $\pm$ 0.017 & 9.93 $\pm$ 0.03 & 10.308 $\pm$ 0.015 & 9.711 $\pm$ 0.020 & (3)\\
    K [mag] & 10.080 $\pm$ 0.016 & 9.877 $\pm$ 0.020 &  10.224 $\pm$ 0.016 & 9.655 $\pm$ 0.018 & (3)\\
    WISE 3.4 $\mu$m [mag] & 10.072 $\pm$ 0.023 & 9.697 $\pm$ 0.022 & 10.154 $\pm$ 0.023 & 9.627 $\pm$ 0.023 & (4)\\
    WISE 4.6 $\mu$m [mag] & 10.091 $\pm$ 0.019 & 9.738 $\pm$ 0.019 & 10.194 $\pm$ 0.02 & 9.663 $\pm$ 0.02 & (4)\\
    WISE 12 $\mu$m [mag] & 10.088 $\pm$ 0.029 & 9.74 $\pm$ 0.06 & 10.12 $\pm$ 0.05 & 9.65 $\pm$ 0.05 & (4)\\
    WISE 22 $\mu$m [mag] & 9.867 & 8.341 & 9.044 & 7.92 & (4)\\
    \hline \hline
    Spectroscopic properties\\
    Spectral type & F & F & F & G & (5)\\
    $T_{eff}$ [K] & 6280 $\pm$ 50 & 6120 $\pm$ 30 & 6110 $\pm$ 50 & 5940 $\pm$ 50 & Sec \ref{Sec:StellarCharacterizationSpectralAnalysis}\\
    log g [cm/s$^2$] & 4.10 $\pm$ 0.08 & 4.15 $\pm$ 0.09 & 4.19 $\pm$ 0.08 & 4.30 $\pm$ 0.08 & Sec \ref{Sec:StellarCharacterizationSpectralAnalysis}\\
    $[$Fe/H$]$ & 0.26 $\pm$ 0.06 & 0.26 $\pm$ 0.06 & 0.28 $\pm$ 0.06 & 0.18 $\pm$ 0.06 & Sec \ref{Sec:StellarCharacterizationSpectralAnalysis}\\
    $v\sin(i)$ [km/s] & 7.8 $\pm$ 0.5 & 7.4 $\pm$ 0.5 & 8.4 $\pm$ 0.5 & 5.3 $\pm$ 0.5 & Sec \ref{Sec:StellarCharacterizationSpectralAnalysis}\\
    \hline \hline
    Bulk properties\\
    Mass [M$_{\odot}$] & 1.38 $\pm$ 0.08 & 1.33 $\pm$ 0.08 & 1.31 $\pm$ 0.08 & 1.16 $\pm$ 0.07 & Sec \ref{Sec:StellarCharacterizationSEDFit}\\
    Radius [R$_{\odot}$] & 1.70 $\pm$ 0.03 & 1.81 $\pm$ 0.06 & 1.51 $\pm$ 0.03 & 1.16 $\pm$ 0.03 & Sec \ref{Sec:StellarCharacterizationSEDFit}\\
    Extinction $A_V$ & 0.03 $\pm$ 0.02 & 0.20 $\pm$ 0.03 & 0.73 $\pm$ 0.05 & 0.06 $\pm$ 0.06 & Sec \ref{Sec:StellarCharacterizationSEDFit}\\
    $L_{\rm bol}$ [L$_{\odot}$] & 3.86 $\pm$ 0.11 & 4.13 $\pm$ 0.21 & 2.89 $\pm$ 0.08 & 1.54 $\pm$ 0.05 & Sec \ref{Sec:StellarCharacterizationSEDFit}\\
    P$_\text{rot}$/sin(i) [d] & 11.0 $\pm$ 0.7 & 12.4 $\pm$ 0.8 & 9.1 $\pm$ 0.5 & 11.1 $\pm$ 1.0 & Sec \ref{Sec:StellarCharacterizationSEDFit}\\
    Age [Gyr] (a) & 2.0 $\pm$ 0.3 & 2.0 $\pm$ 0.4 & 0.3 $\pm$ 0.1 & 1.3 $\pm$ 0.2 & Sec \ref{Sec:StellarCharacterizationSEDFit}\\
    \hline \hline
    \end{tabular}
    \label{tab:StellarParameters}
    }
    \tablebib{(1) \cite{2020Gaia}; (2) ExoFOP website; (3) \cite{20032MASS}; (4) \cite{2010WISE}; (5) Table 5 \cite{Pecaut2013}; (a) age obtained from $v\sin(i)$}
\end{table*}

\section{Analysis} \label{Sec:Analysis}

\subsection{Joint fit of the photometric and spectroscopic data with {\tt juliet}} \label{Sec:AnalysisJointFit}

To obtain values for the parameters of the planetary systems of the four targets, we performed a joint fit of the photometric and spectroscopic data using the Python package {\tt juliet} \citep{juliet2019}.
Here, the two datasets are the photometric data (time, flux, and error of the flux) and the calculated radial velocities (RVs) derived from the spectroscopic observations (time, RV, and error of the RV). The program assumes that the two datasets are independent from each other, but some parameters (e.g., the period or the mid-transit time) can be fitted from both sets. By fitting all the data simultaneously, the common parameters can be  better constrained.
\\
\\
We fit for the following transit parameters: Orbital period (P), transit mid-point/time of transit-center ($t_0$), planet-to-star-radius ratio (Rp/Rs: p), impact parameter (b), eccentricity (e), argument of periastron ($\omega$), and the stellar density ($\rho$).
Additional parameters from the instruments are the dilution factor ($D_{i}$) and the mean out-of-transit flux/relative flux ($M_{i}$) of each instrument, together with a jitter term ($\sigma_{w,i}$) that is added in quadrature to the errorbars of the instruments. Furthermore, the limb-darkening is fitted with two parameters ($q_{1,i}$, $q_{2,i}$) for each instrument.
To fit the radial velocities the following parameters are used: orbital period, transit midpoint, radial velocity semi-amplitude (K), eccentricity, and argument of periastron.
Additionally, for each instrument, the median systemic velocity ($\mu_i$) and a corresponding jitter ($\sigma_{w,i}$) term were fitted.
The parameters with their initial priors and the posterior values from the joint photometric and spectroscopic fit are shown in tables online (see Sect. \ref{Sec:DataAvailability}) for the four targets (TOI-1295, TOI-2580, TOI-6016, and TOI-6130).
\\
The targets we observe in this study are hot Jupiter candidates with short periods (P $<$ 5~days). They were initially reported to be TOIs, based on observations with the TESS satellite, and were all observed throughout at least one sector (duration $\sim$ 27~days). Hence, the period and transit midpoint were already well-constrained from the TESS observations. Thus, we have decided to use normal priors centered around the predicted values for the orbital period and the transit mid-point.
For the planet-to-star-radius ratio and the impact parameter, we have allowed for all physically possible values in a uniform distribution to avoid biasing the fit.
\\
The eccentricity and the argument of periastron are difficult to constrain accurately from the fit. In particular, the eccentricity can be biased toward non-zero values in the fit \citep{Lucy1971}. Therefore, we performed two fitting runs for each target. Between those, we only changed the priors of the eccentricity and longitude of periastron. In the first run, we set uniform priors allowing values of the eccentricity to range between 0 and 1, as well as for the longitude of periastron [0$^\circ$, 180$^\circ$]; for the second run, we fixed them to e = 0, and $\omega$ = 90$^\circ$. When deciding on the best results, we inspected the values of the Bayesian evidence ($\ln Z$).
\\
We have selected the preferred model by taking the model with the higher Bayesian evidence value. Here, a difference of $ \Delta \ln Z >$ 5 constitutes a strong preference for one model over the other, while a $ \Delta \ln Z >$ 2 indicates a weak preference \citep{trotta2008bayes}.
The stellar density can be better constrained from literature values and our own separate analysis of the spectroscopic observations than with the here performed fit. Thus, a normal prior around a value from the literature data was chosen. 
\\
For the limb darkening parameters, we  chose a normal prior for the TESS data and, for the other instruments, we decided to use uniform priors over the complete range of possible values.
The mean-out-of-transit flux is expected to be around 1 for the normalized data. Thus, a normal prior was chosen here.
For the jitter term, a log uniform prior is chosen to allow a wide range of values.
The dilution factor ($D_i$) for each photometric dataset is fixed to one because the SPOC data is corrected for contamination of nearby stars.
For the systematic RV offset a uniform prior was  chosen to ensure we could avoid bias.
Similarly, a uniform prior was  chosen for the semi-amplitude of the RV curve, allowing for a wide range of reasonable values.
For the jitter term of the RV curve, we again chose a loguniform prior.


\subsection{Parameter values from the fit} \label{Sec:ParResults}

\subsubsection*{TOI-1295 b}

When performing a joint {\tt juliet} fit of the photometric and RV data, the model for the RV curve does not include the velocity deviations caused by the RM-effect during transit. Thus, we have decided to exclude all observations (in total 37 observations) that happened at least partially during the transit times.
Between the circular and eccentric {\tt juliet} models, the eccentric model was favored for TOI-1295 b, with a difference seen in the log-evidence values of $\sim$ 4.6 constituting a moderate-to-strong preference. However, the derived eccentricity is rather small at e = 0.03 $\pm$ 0.02.
\\
 With TOI-1295, we aimed to move toward observing fainter stars (in the V-band a magnitude $\geq$ 11) that have been observed with the MaHPS before. Using the RV method, we found that for TOI-1295 with a V-band magnitude of 11.3 mag a mass precision of better than 15$\%$ can be achieved (planetary mass: 1.5 $\pm$ 0.2~$\text{M}_{Jup}$).
The resulting fits to the TESS data and the ground-based photometric observations are shown in a figure published online (see Sect. \ref{Sec:DataAvailability}) and the RV fit is shown in Fig. \ref{fig:rvplot}. The parameter values (priors and posteriors) obtained from this fitting process are shown in an  online table (see Sect. \ref{Sec:DataAvailability}). 
\\

\subsubsection*{TOI-2580 b}

\begin{figure*}[htp]
    \centering
    \begin{subfigure}{.48\textwidth}
        \centering
        TOI-1295
        \includegraphics[width=\linewidth]{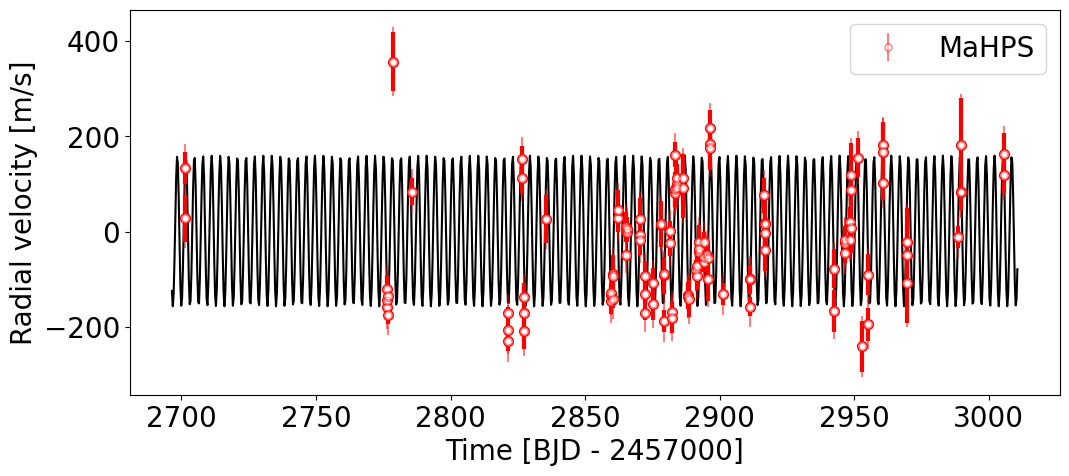}%
    \end{subfigure}\hfill
    \centering
    \begin{subfigure}{.48\textwidth}
        \centering
        TOI-2580
        \includegraphics[width=\linewidth]{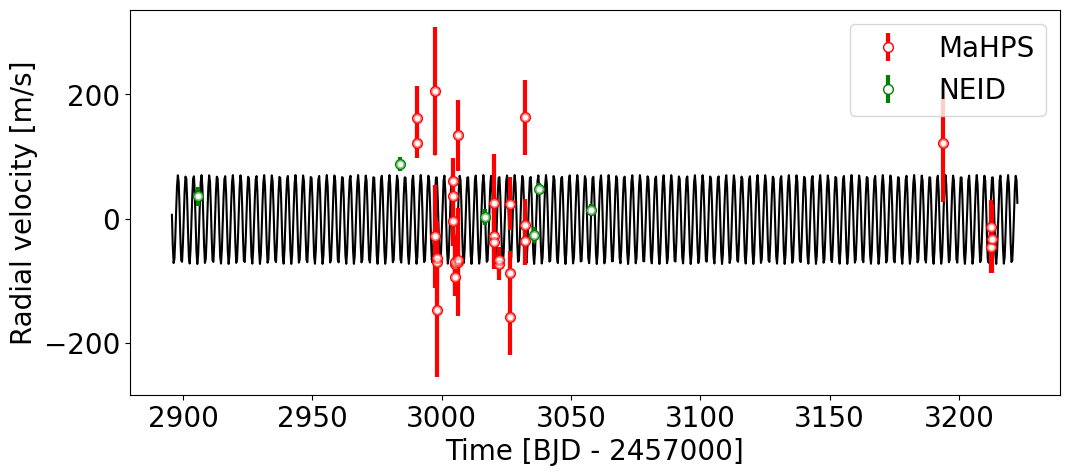}
    \end{subfigure}
    \hfill
    \centering
    \begin{subfigure}{.48\textwidth}
        \centering
        \includegraphics[width=\linewidth]{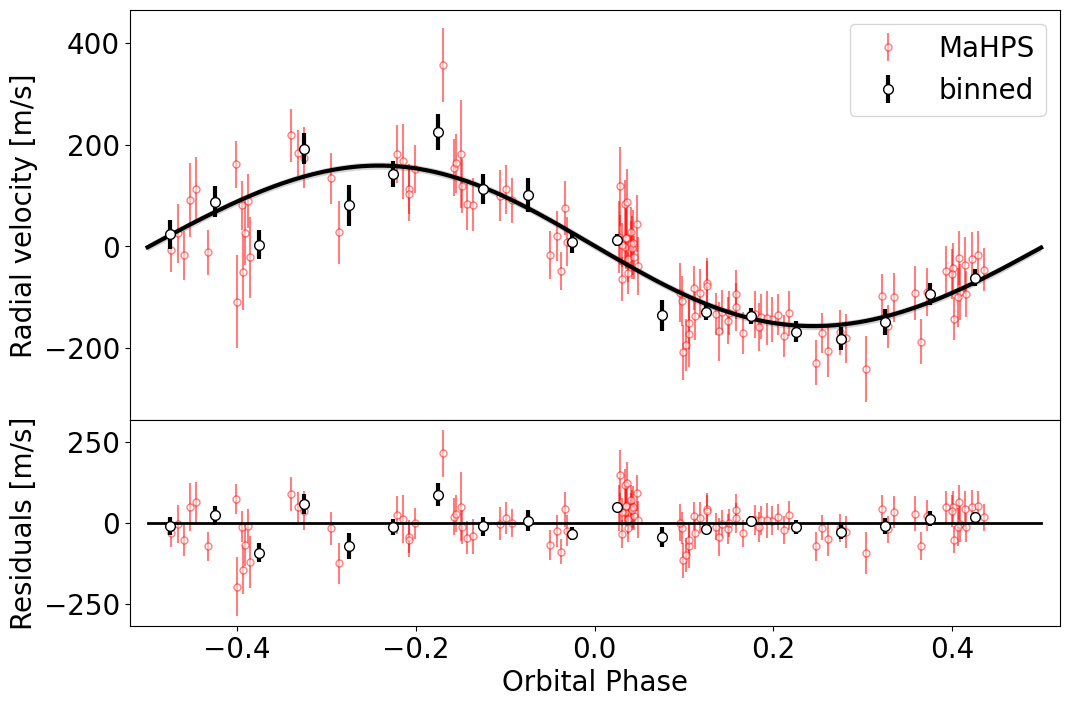}
     \end{subfigure}\hfill
    \centering
    \begin{subfigure}{.48\textwidth}
        \centering
        \includegraphics[width=\linewidth]{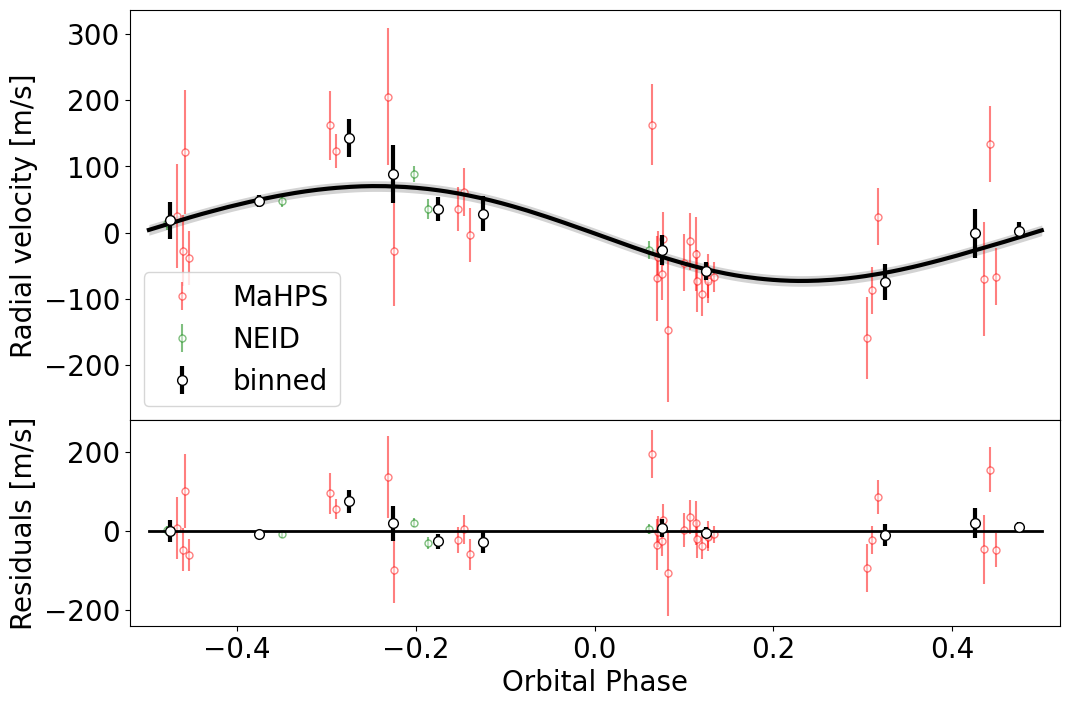}
    \end{subfigure}
    \centering
    \begin{subfigure}{.48\textwidth}
        \centering
        TOI-6016
        \includegraphics[width=\linewidth]{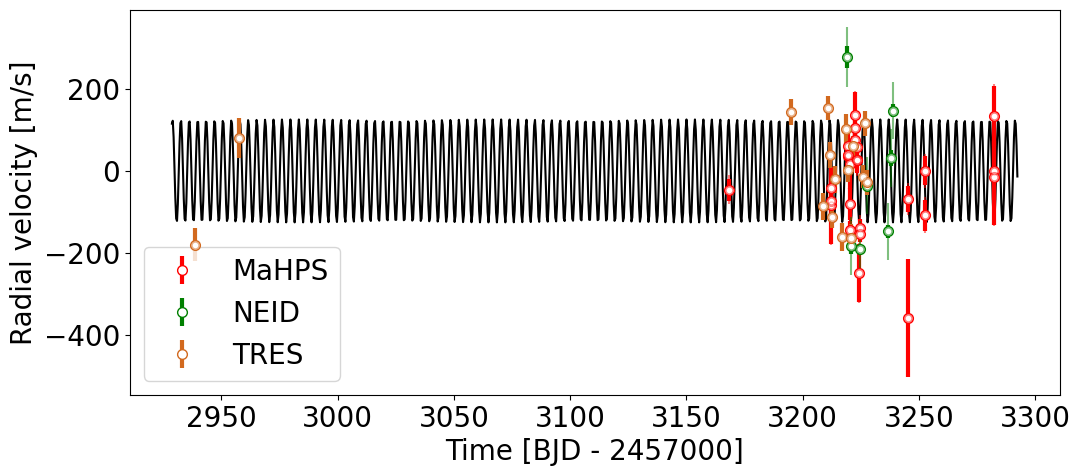}%
    \end{subfigure}\hfill
    \centering
    \begin{subfigure}{.48\textwidth}
        \centering
        TOI-6130
        \includegraphics[width=\linewidth]{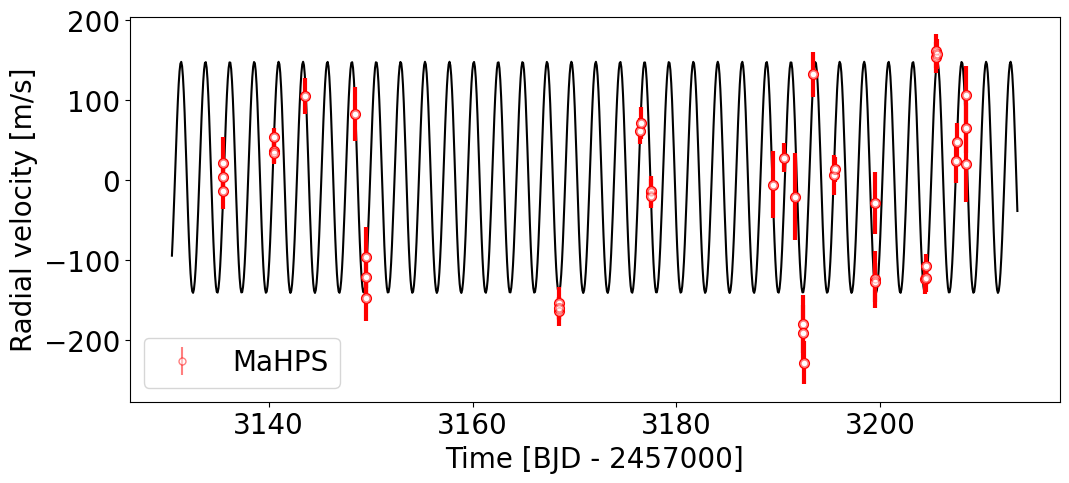}
    \end{subfigure}
    \hfill
    \centering
    \begin{subfigure}{.48\textwidth}
        \centering
        \includegraphics[width=\linewidth]{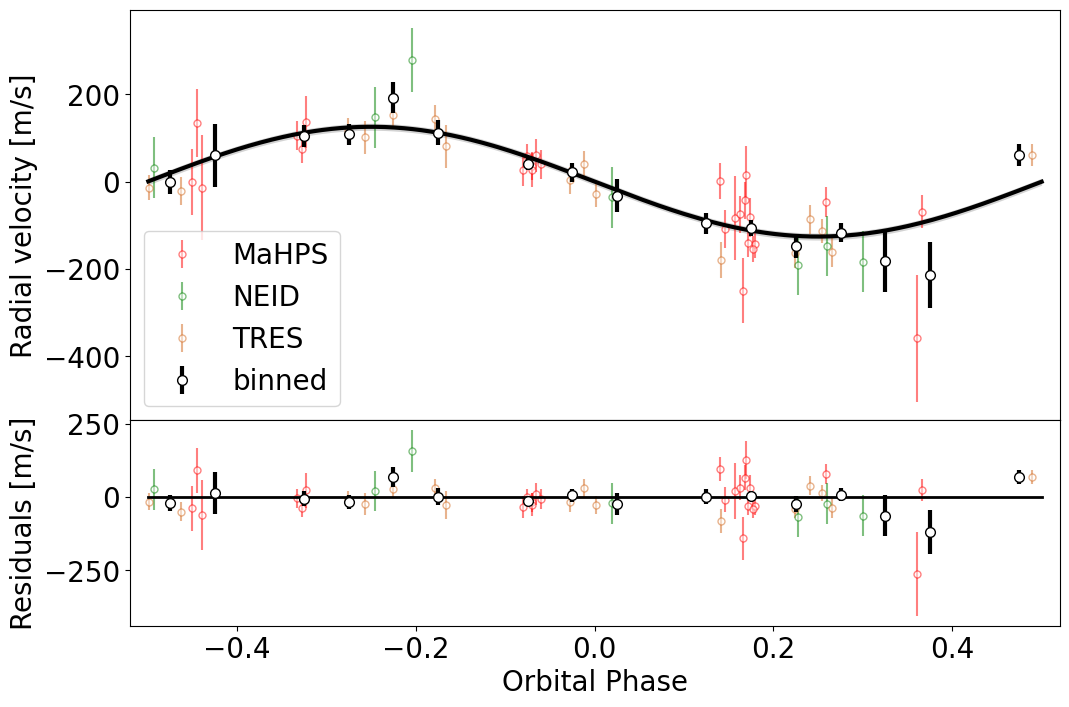}
     \end{subfigure}\hfill
    \centering
    \begin{subfigure}{.48\textwidth}
        \centering
        \includegraphics[width=\linewidth]{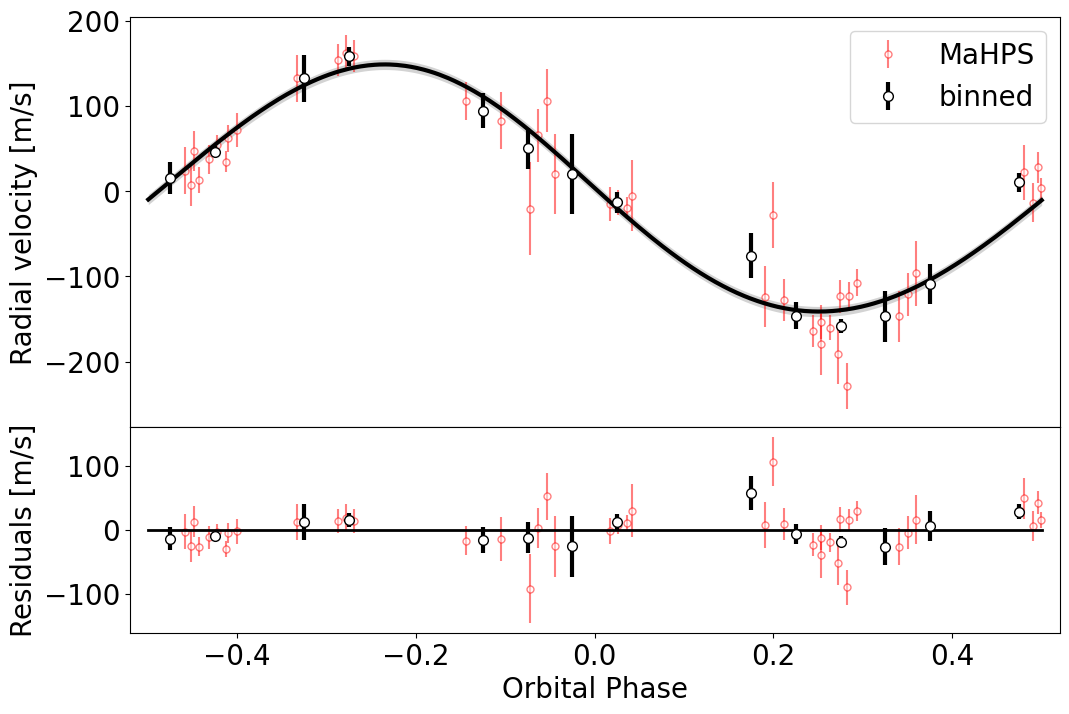}
    \end{subfigure}
    \caption{{\tt juliet} Fit of the radial velocities of TOI-1295 (upper left), TOI-2580 (upper right), TOI-6016 (lower left), and TOI-6130 (lower right) obtained with the MaHPS (red), TRES (brown), and NEID (green). For TOI-1295, the data taken during the transit for the measuring of the RM effect were excluded from the fit. The gray area around the folded rv-curve in the middle panel shows the RV error of the joint fit}
    \label{fig:rvplot}
\end{figure*}

For TOI-2580, the eccentric model was favored (difference in the log-evidence values $\sim$ 3.3). We derived a small orbital eccentricity of e = 0.07 $\pm$ 0.06. The resulting fits to the TESS data and the ground-based photometric observations are shown in a figure online (see Sect. \ref{Sec:DataAvailability}). The resulting fit to the radial velocity data is shown in Fig. \ref{fig:rvplot}. The parameter values (priors and posteriors) obtained from this fitting process are shown in a table online (see Sect. \ref{Sec:DataAvailability}).


\subsubsection*{TOI-6016 b}

For TOI-6016 the difference between the circular and the eccentric model was smaller ($\sim$ 1), however, we have chosen the circular model as the latter is slightly favored. The chosen model does not change the results significantly, as the eccentrity value for the uniform model was small (e = 0.19 $\pm$ 0.06). The figure of the photometric fit is shown online (see Sect. \ref{Sec:DataAvailability}) and \ref{fig:rvplot} displays the resulting fit to the RV data. The parameter values (priors and posteriors) obtained from this fitting process are shown in a table online (see Sect. \ref{Sec:DataAvailability}).


\subsubsection*{TOI-6130 b}

For TOI-6130, the eccentric model was slightly favored as the difference in the log-evidence levels between the uniform and fixed was $\sim$ 2.3. We derived a small orbital eccentricity of  e = 0.045 $\pm$ 0.030. The resulting fits to the TESS data and the ground-based photometric observations are shown in an  online figure (see Sect. \ref{Sec:DataAvailability}). The resulting fits to the radial velocity data are shown in Fig. \ref{fig:rvplot}.
The parameter values (priors and posteriors) obtained from this fitting process are shown in an  online table (see Sect. \ref{Sec:DataAvailability}).


\subsection{Search for companion planets}

An important tracer for the formation mechanisms of hot Jupiters is the architecture of the planetary systems and the properties of their companions. 
To investigate the formation mechanism responsible for the four presented hot Jupiters, we used the photometric data from TESS and our RV data to search for more planets in the four systems.

\subsubsection{RV trend and periodicity in RV curve}

To search for signals of additional planets in the RV data, we used the Python package RVSearch \citep{rvsearch2021}. This program searches for periodicities in the RV data by constructing $\Delta$BIC goodness-of-fit periodograms from a comparison of the fit of a single-planet Keplerian model from RadVel \citep{fulton2018radvel} to a model without a planet over a grid of periods.
If a signal exceeding a given false-alarm probability (FAP) threshold is found, a two-planet model is compared to the one-planet model in another periodogram analysis, until no significant signals are left in the data.
\\
For our analysis, we set the minimum period to one day and the maximum to three times our observational baseline. The FAP limit for a detection was set to $1 \%$. While RVSearch can fit offsets in the RV data between different instruments, we chose to use the offsets derived from the {\tt juliet} fits to correct the different RVs to a common zero-point. This was done to ensure that the ability to recover signals is not limited by how well the offsets between different instruments can be determined. For all four planets, in the one-planet case, the highest peak in the periodogram was located at the period of the confirmed planets (left column in Fig. \ref{fig:RVsearch}). However, for TOI-2580, we were not able to recover the planetary signal at the given FAP. For the other three stars, we continued to search for further signals, taking into account the presence of the first planet. For TOI-1295 and TOI-6130, we did see secondary peaks, which were also above 0.01 FAP; however, these disappeared after we subtracted the signal of the detected planet, indicating that those were, in fact, the aliases of the first planet period.
\\
As shown in the middle column of Fig. \ref{fig:RVsearch}, no significant secondary signals were found in any of the systems. For TOI-1295 there is a distinct peak at 144~days with the highest power in the periodogram, but it is far below the $1 \%$ FAP limit. These results suggest that there are no detectable companions to our confirmed hot Jupiters. 
\\
However, to further investigate the significance of these findings, we used the injection recovery test implemented in RVSearch to check whether we would be able to detect potential other planets in our RV data. We drew synthetic planets from a log-uniform period and M $\sin(i)$ distribution and injected them into our data. RVSearch was then run to try to recover those signals and construct a completeness distribution for the dataset. We injected 5000 signals with periods between 1 and 5000~days and RV amplitudes between 1 and 1000 m/s into the three datasets where we were able to recover the signal of the TESS planet. The resulting completeness plots are shown in the right column of Fig. \ref{fig:RVsearch}.
The blue dots show the planets that could be recovered from the program; whereas this was not possible for the planets marked with red circles. The background color describes the fraction of recovered planetary companions where the black line is the $50 \%$ threshold.
The black dashed line is the brown dwarf mass which separates them from the planetary candidates. Here, we use the deuterium burning limit of about 13 Jupiter masses as the separation between exoplanets and brown dwarfs. The black crosses show the here confirmed planets of the targets.
It becomes apparent from these plots that the available data only allow for the detection of a limited range of parameters of planetary companions. Less massive planets, starting from below 1 Jupiter mass (displayed as the lower black dashed line in the plots), can mainly be detected up to a period of $\sim$ 10~days. Longer period companions can only be found with a maximum orbital duration of $\sim$ 300~days and only with masses higher than 1 Jupiter mass. Thus, with the available data, finding  long-period massive companions that would be expected from high-eccentricity formation theories is unlikely. To determine the presence of such companions, observations with a longer baseline are needed. Nevertheless, we can at least exclude, with some degree of confidence, the presence of nearby Jupiter-sized planetary companions.

\subsubsection{Transit-timing variations}
As a secondary step to search for additional planets in the system, we checked the photometric data from TESS for signs of transit-timing variations (TTVs) using {\tt juliet}. For this purpose, we  fit every transit midpoint ($T_n$) individually to find periodically occurring deviations from the predicted timing of these midpoints that could be caused by the gravitational pull of a nearby planetary companion. For the three planets, TOI-2580 b, TOI-6016 b, and TOI-6130 b, we used the data from all available sectors. Due to the large number of observations for TOI-1295 b, we only used sectors 47-60 (excluding 58 where the star was not observed) for the TTV analysis.
\\
The $T_n$ values were fitted as parameters with {\tt juliet}. The resulting O-C-diagrams (figure available online; see Sect. \ref{Sec:DataAvailability}) that show the difference between the observed and computed transit midpoints in the TESS sectors show only small deviations (less than approximately 4 min, 5 min, 5 min, and 1.5 min for TOI-1295, TOI-2580, TOI-6016, and TOI-6130, respectively). Thus, we did not find any signs of planetary companions for any of the four targets.
\\
To further investigate these results, we created Lomb-Scargle diagrams from the derived transit midpoints to search for signs of periodicity. These diagrams were  created using the Lomb-Scargle periodogram from {\tt astropy} \citep{astropy:2013, astropy:2018, astropy:2022} and the corresponding false alarm levels were calculated using boostrapping. We find no peaks exceeding a FAP of $10 \%$ for any of the targets (see Fig. \ref{fig:LS}).
\\
The baseline of the TESS observations is too short to allow us to draw any definitive conclusions. As illustrated in \cite{Wu2023}, the first signs of TTVs in hot Jupiter systems were found only after including the full Kepler baseline. Those signals were not detectable when using only six quarters of the Kepler data \citep{Steffen2012}. As  six quarters of Kepler data already offer a much larger number of transits than the TESS data of our systems, we would  not be likely to detect typical TTVs, even if they were present in these systems.


\section{Discussion} \label{Sec:Discussion}

\subsection{Planetary properties} \label{Sec:planet_porporties}

\begin{figure}[htp]
    \centering
    \includegraphics[width=\linewidth]{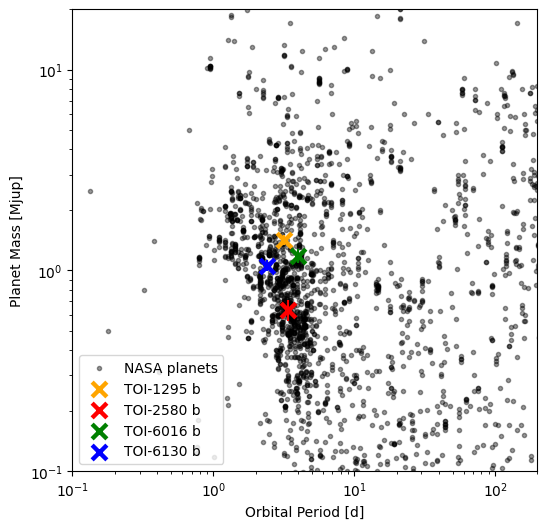}
    \caption{Period-mass plot of the four confirmed exoplanets showing them as hot Jupiters}
    \label{fig:HotJupiters_PerMpl}
\end{figure}

\begin{figure}[htp]
    \centering
    \includegraphics[width=\linewidth]{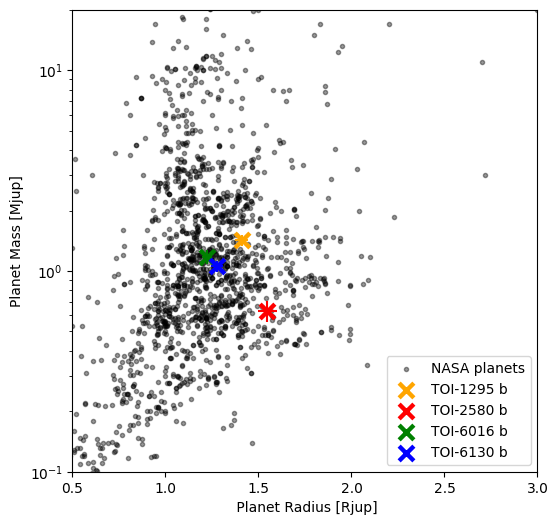}
    \caption{Mass-radius plot of the four confirmed exoplanets displaying TOI-2580 b with the lowest density}
    \label{fig:HotJupiters_MassRadius}
\end{figure}

Using the stellar parameters from Sect. \ref{Sec:StellarCaracterization}, we calculated the physical parameters of the planets. The most important parameters are shown in Table \ref{tab:PhysicalParameters}.
\begin{table*}[htp]
    \centering
    \caption{Summary of the key parameters of the four targets, TOI-1295 b, TOI-2580 b, TOI-6016 b, and TOI-6130 b, in physical units}
    \begin{tabular}{l|c|c|c|c}
     Parameter & TOI-1295 & TOI-2580 & TOI-6016 & TOI-6130 \\
     \hline \hline
     P$_{pl}$ [d] & 3.1968838 $\pm$ 0.0000005 & 3.397750 $\pm$ 0.000002 & 4.023687 $\pm$ 0.000003 & 2.392679 $\pm$ 0.000002\\
     a [AU] & 0.047 $\pm$ 0.002 & 0.048 $\pm$ 0.003 & 0.055 $\pm$ 0.002 & 0.036 $\pm$ 0.002\\
     M$_{pl}$ [M$_{Jup}$] & 1.42 $\pm$ 0.08 & 0.63 $\pm$ 0.08 & 1.17 $\pm$ 0.09 & 1.05 $\pm$ 0.06 \\
     R$_{pl}$ [R$_{Jup}$] & 1.40 $\pm$ 0.08 & 1.55 $\pm$ 0.05 & 1.22 $\pm$ 0.03 & 1.28 $\pm$ 0.03 \\
     $\rho$ [$\text{g}/\text{cm}^3$] & 0.65 $\pm$ 0.05 & 0.22 $\pm$ 0.04 & 0.81 $\pm$ 0.08 & 0.64 $\pm$ 0.06\\
     \hline
     t0 [d] & 2459913.37999$\pm$0.00020 & 2458839.4534$\pm$0.0005 & 2459877.7930$\pm$0.0003 & 2459849.63919$\pm$0.00016\\
     K [m/s] & 158$\pm$7 & 70$\pm$8 & 125$\pm$8 & 145$\pm$6\\ 
     e & 0.024$\pm$0.020 & 0.08$\pm$0.04 & fixed: 0.0 & 0.036$\pm$0.018\\
     $\omega$ [$^\circ$] & 80$\pm$40 & 114$\pm$30 &  fixed: 90.0 & 42$\pm$30\\
     T$_{eq}$ [K] & 2360$\pm$50 & 2410$\pm$60 & 1890$\pm$40 & 1750$\pm$40\\
     Instellation [I$_{\oplus}$] & (1.80$\pm$0.13)$\times 10^3$ & (1.73$\pm$0.13)$\times 10^3$ & (9.4$\pm$0.5)$\times 10^2$ & (1.15$\pm$0.08)$\times 10^3$ \\
     Transit duration [min] & 370$\pm$15 & 503$\pm$25 & 342$\pm$13 & 145$\pm$5\\
    \end{tabular}
    \label{tab:PhysicalParameters}
\end{table*}
\\
Figure \ref{fig:HotJupiters_PerMpl} shows the position of the four planets in the period-mass-space of known exoplanets from the NASA Exoplanet Archive. All four of our targets are located in the well-populated part of the hot Jupiter parameter space. The three targets, TOI-1295 b, TOI-6016 b, and TOI-6130 b, are likely to be similar in terms of their bulk composition, with all of them having densities close to that of Saturn. In contrast, TOI-2580 b seems to be more inflated, having the largest radius of the four planets with only half the mass of the others resulting in a significantly lower bulk density (see Fig. \ref{fig:HotJupiters_MassRadius}).
\\
Regarding the eccentricities, we find that for TOI-6016 b the circular orbit with zero eccentricity was the preferred model. While for the other three the eccentric model was favored, all derived eccentricity values are very low, at 0.03 $\pm$ 0.02 (TOI-1295 b), 0.07 $\pm$ 0.06 (TOI-2580 b), and 0.045 $\pm$ 0.030 (TOI-6130 b). This is expected as most hot Jupiters should have their orbits circularized from the strong tidal forces of their host star \citep{Correia2010}.
\\

\subsection{Radius inflation}
\begin{figure}[htp]
    \centering
    \includegraphics[width=\linewidth]{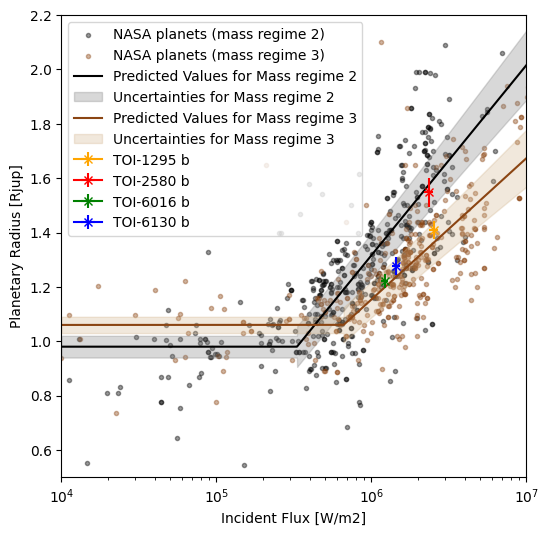}
    \caption{Radius inflation. Black data points show the confirmed planets from the NASA exoplanet archive in mass regime 2 (0.37 - 0.98~M$_{Jup}$). Brown data points show the confirmed planets from the NASA exoplanet archive in mass regime 3 (0.98 - 2.50~M$_{Jup}$). All four targets agree perfectly with the computed radius relations developed by \cite{Sestovic2018}, displayed as black and brown lines with error bars in the corresponding mass regimes.}
    \label{fig:RadiusInflation}
\end{figure}

As mentioned earlier in this work (Sect. \ref{Sec:planet_porporties}), TOI-2580 b seems to be significantly inflated compared to the other three planets.
One of the current formalisms to describe the radius inflation of planetary candidates has been developed by \cite{Sestovic2018} by applying a hierarchical Bayesian model to a sample of 286 gas giants. \cite{Sestovic2018} used the observational data of previously studies hot Jupiters to design a purely empirical model that captures the radii of these exoplantes.
In the following, we  describe how we checked whether the radii of our four new planets are in agreement with this proposed model. Initially, we computed the expected radius based on this model, so it can be compared with the fitted radii. 
\\
Previous works investigating the radius inflation of hot Jupiters has found that this occurs above a certain limit, $F_s$, of the incident flux, F, on the planet \citep[e.g.,  ][]{Demory2011}.
This is expressed as Eq. (1) in \cite{Sestovic2018}:
\begin{equation}
\frac{R}{R_{Jup}} =
\begin{cases}
  C, & F < F_s, \\
  C + A \cdot (log F - log F_s), &  F \geq F_s. \\
\end{cases}
\label{equ:InflR}
\end{equation}
The study from \cite{Sestovic2018}  found that  parameters A, C, and F$_s,  $    used to compute the radii of hot Jupiters (see Eq. \ref{equ:InflR}), vary with planetary mass.
The authors  split the planetary mass ranges into four categories; namely,  exoplanets with masses of $<$M$_{Jup}$ (sub-Saturns), those with 0.37~$<$~M$_{Pl}$~$<$~0.98~M$_{Jup}$ (sub-Jupiters), and planets with larger masses than Jupiter, with 0.98~$<$~M$_{Pl}$~$<$~2.5~M$_{Jup}$ and M$_{Pl}$~$>$~2.5~M$_{Jup}$.
This is especially interesting, as one of our objects, TOI-2580 b (with the lowest mass of the four targets) falls into their second mass regime (0.37~-~0.98~M$_{Jup}$), while the other three planets, TOI-1295 b, TOI-6016 b, TOI-6130 b, reside in the third mass regime (0.98~-~2.50~M$_{Jup}$).
The fluxes above which the radius inflation is expected to occur are $log_{10} F_s > 5.52 \genfrac{}{}{0pt}{}{\scriptstyle +0.07}{\scriptstyle -0.09}$ for TOI-2580 b, and $log_{10} F_s > 5.82 \genfrac{}{}{0pt}{}{\scriptstyle +0.09}{\scriptstyle -0.11}$ for the other three objects.
\\
The incident flux on the exoplanet surface has been computed using Eq. (6) from \cite{Sestovic2018}:
\begin{equation}
    F = \frac{R_*^2}{a^2} \sigma T_*^4
    \label{equ:IncidentFlux}
.\end{equation}
\\
We inserted the values for $R_*$, a, and $T_*$. Specifically, we  used $R_*$ from the SED fitting (see Table \ref{tab:StellarParameters}), a from the joint fit (see Table \ref{tab:PhysicalParameters}), and for $T_*$ we have used the weighted average values from $T_{eff}$ (see Tables \ref{tab*:ComparisonStellarParameters1}, \ref{tab*:ComparisonStellarParameters2}), which resulted in the following incident flux values (Table \ref{tab:IncidentFluxes}):
\\
\begin{table}[htp]
    \centering
    \caption{Incident stellar fluxes on the four targets}
    \begin{tabular}{c|l|l}
         & F [W/m$^2$] & log$_{10}$ F\\
    \hline \hline
    TOI-1295 b & (2.45 $\pm$ 0.18)$\times 10^6$ & 6.39 $\pm$ 0.04\\
    TOI-2580 b & (2.35 $\pm$ 0.19)$\times 10^6$ & 6.37 $\pm$ 0.04\\
    TOI-6016 b & (1.28 $\pm$ 0.08)$\times 10^6$ & 6.107 $\pm$ 0.025\\
    TOI-6130 b & (1.57 $\pm$ 0.11)$\times 10^6$ & 6.195 $\pm$ 0.029\\
    \end{tabular}
    \label{tab:IncidentFluxes}
\end{table}
\\
Thus, even when considering the errors on the flux limits and on the computed properties, the incident flux values on all four planets lie above $F_s$ for the respective mass.
So, based on the values presented in the paper, the following equations have been used to compute the predicted planetary radii \citep[see Table 1 of ][]{Sestovic2018}.
For TOI-2580 b, with the above-stated values for $F_s$, and the obtained computed value for F, we have:
\begin{equation}
    \frac{R_{pl}}{R_{Jup}} = (0.98 \pm 0.04) + \bigg( 0.70^{+0.07}_{-0.0} \bigg) \cdot (log F - log F_s) 
.\end{equation}
For TOI-1295 b, TOI-6016 b, TOI-6130 b the formula is:
\begin{equation}
    \frac{R_{pl}}{R_{Jup}} = (1.06 \pm 0.03) + (0.52 \pm 0.07) \cdot (log F - log F_s)  
.\end{equation}
When comparing the radii computed using the parameterization from this model, $R_{model}$, and the fitted radii $R_{fit}$ shows a good agreement within the errorbars (see Table \ref{tab:Radii} and Fig. \ref{fig:RadiusInflation}).
Additionally, it becomes clear that all four exoplanets are inflated.
\\
\begin{table}[htp]
    \centering
     \caption{Comparison of the computed radii from the model and the radii found from the joint fit of the photometric and RV data}
    \begin{tabular}{c|l|l}
         & $R_{model}$ [$R_{Jup}$] & $R_{fit}$ [$R_{Jup}$]\\
    \hline \hline
    TOI-1295 b & 1.37 $\pm$ 0.02 & 1.41 $\pm$ 0.04\\
    TOI-2580 b & 1.59 $\pm$ 0.02 & 1.55 $\pm$ 0.05\\
    TOI-6016 b & 1.22 $\pm$ 0.01 & 1.23 $\pm$ 0.03\\
    TOI-6130 b & 1.27 $\pm$ 0.02 & 1.28 $\pm$ 0.04\\
    \end{tabular}
    \label{tab:Radii}
\end{table}
\\
\\
It becomes clear that both the incident flux on the exoplanet and also its mass determine the radius size. Even though TOI-2580 b receives similar flux as the other three objects, its mass is significantly smaller leading to the inference of a much more inflated planet.

\section{Summary} \label{Sec:Conclusions}

In this paper, we have confirmed and analyzed the four TOI planet candidates: TOI-1295 b, TOI-2580 b, TOI-6016 b, and TOI-6130 b. We used joint fits of the RV data from spectroscopic observations and photometric observations of the transit light curves from the TESS satellite and ground-based observatories. We  confirmed their planetary nature and measured their precise radii and masses. 
\\
All four hot Jupiters show low eccentricity values and no signs of nearby companions. These findings would support high-eccentricity as the likely formation mechanism for those four hot Jupiters. However, a long-period perturber, which is an essential prerequisite for high-eccentricity migration,  was not found. A longer baseline of RV observations is needed to draw definitive conclusions on the formation of these object.
The inflated radii of the four hot Jupiters are in agreement with the model predicted by \cite{Sestovic2018}, thus supporting that the radius of hot Jupiters depends mainly on the incident flux from its host star and the planetary mass. Lastly, we show that the MaHPS spectrograph even with only a 2.1~m telescope is able to detect hot Jupiters around relatively faint stars (V~$\sim$ 12~mag) with a good level of precision.

\section{Data availability} \label{Sec:DataAvailability}

The fit of the photometric light curves,  tables containing the prior and posterior values from our joint fit of the photometric and RV data,  plots displaying the S/N of the MaHPS observations, and  O-C diagrams can be found online at \url{https://zenodo.org/records/13840492}

\begin{acknowledgements}

The Wendelstein 2.1~m telescope project was funded by the Bavarian government and by the German Federal government through a common funding process. Part of the 2.1 m instrumentation including some of the upgrades for the infrastructure were funded by the Cluster of Excellence “Origin of the Universe” of the German Science foundation DFG.


This paper contains data taken with the NEID instrument, which was funded by the NASA-NSF Exoplanet Observational Research (NN-EXPLORE) partnership and built by Pennsylvania State University. NEID is installed on the WIYN telescope, which is operated by the National Optical Astronomy Observatory, and the NEID archive is operated by the NASA Exoplanet Science Institute at the California Institute of Technology. NN-EXPLORE is managed by the Jet Propulsion Laboratory, California Institute of Technology under contract with the National Aeronautics and Space Administration.
\\
Data presented herein were obtained at the WIYN Observatory from telescope time allocated to NN-EXPLORE through the scientific partnership of the National Aeronautics and Space Administration, the National Science Foundation, and NOIRLab.
\\
This work was supported by NASA WIYN PI Data Awards (2022B-963260, 2023A-669281, PI: Yee), administered by the NASA Exoplanet Science Institute.
The authors are honored to be permitted to conduct astronomical research on Iolkam Du’ag (Kitt Peak), a mountain with particular significance to the Tohono O’odham. 


We acknowledge the use of public TESS data from pipelines at the TESS Science Office and at the TESS Science Processing Operations Center.
\\
Resources supporting this work were provided by the NASA High-End Computing (HEC) Program through the NASA Advanced Supercomputing (NAS) Division at Ames Research Center for the production of the SPOC data products.
\\
This paper made use of data collected by the TESS mission and are publicly available from the Mikulski Archive for Space Telescopes (MAST) operated by the Space Telescope Science Institute (STScI). Funding for the TESS mission is provided by NASA’s Science Mission Directorate.


This work makes use of observations from the LCOGT network. This paper is based on observations made with the Las Cumbres Observatory’s education network telescopes that were upgraded through generous support from the Gordon and Betty Moore Foundation.\\


This research has made use of the Exoplanet Follow-up Observation Program (ExoFOP; DOI: 10.26134/ExoFOP5) website, which is operated by the California Institute of Technology, under contract with the National Aeronautics and Space Administration under the Exoplanet Exploration Program.\\


Funding for the TESS mission is provided by NASA's Science Mission Directorate. KAC and CNW acknowledge support from the TESS mission via subaward s3449 from MIT.\\



This article is based on observations made with the MuSCAT2 instrument, developed by ABC, at Telescopio Carlos Sánchez operated on the island of Tenerife by the IAC in the Spanish Observatorio del Teide. This work is partly supported by JSPS KAKENHI Grant Numbers JP18H01265 and JP18H05439, and JST PRESTO Grant Number JPMJPR1775. We acknowledge financial support from the Agencia Estatal de Investigaci\'on of the Ministerio de Ciencia e Innovaci\'on MCIN/AEI/10.13039/501100011033 and the ERDF "A way of making Europe" through project PID2021-125627OB-C32, and from the Centre of Excellence "Severo Ochoa" award to the Instituto de Astrofisica de Canarias.

This work is partly supported by JSPS KAKENHI Grant Number JP24H00017 and JSPS Bilateral Program Number JPJSBP120249910.\\


This research has made use of the NASA Exoplanet Archive, which is operated by the California Institute of Technology, under contract with the National Aeronautics and Space Administration under the Exoplanet Exploration Program.\\


Adam Popowicz was financed by grants 02/140/RGJ24/0031 and BK-250/RAu-11/2024.

\end{acknowledgements}

\bibliographystyle{aa}
\bibliography{bib}

\onecolumn

\begin{appendix}

\section{Search for companion planets}

\begin{figure*}[htp]
    \centering
    \begin{minipage}{\textwidth}
        \includegraphics[width=0.31 \textwidth]{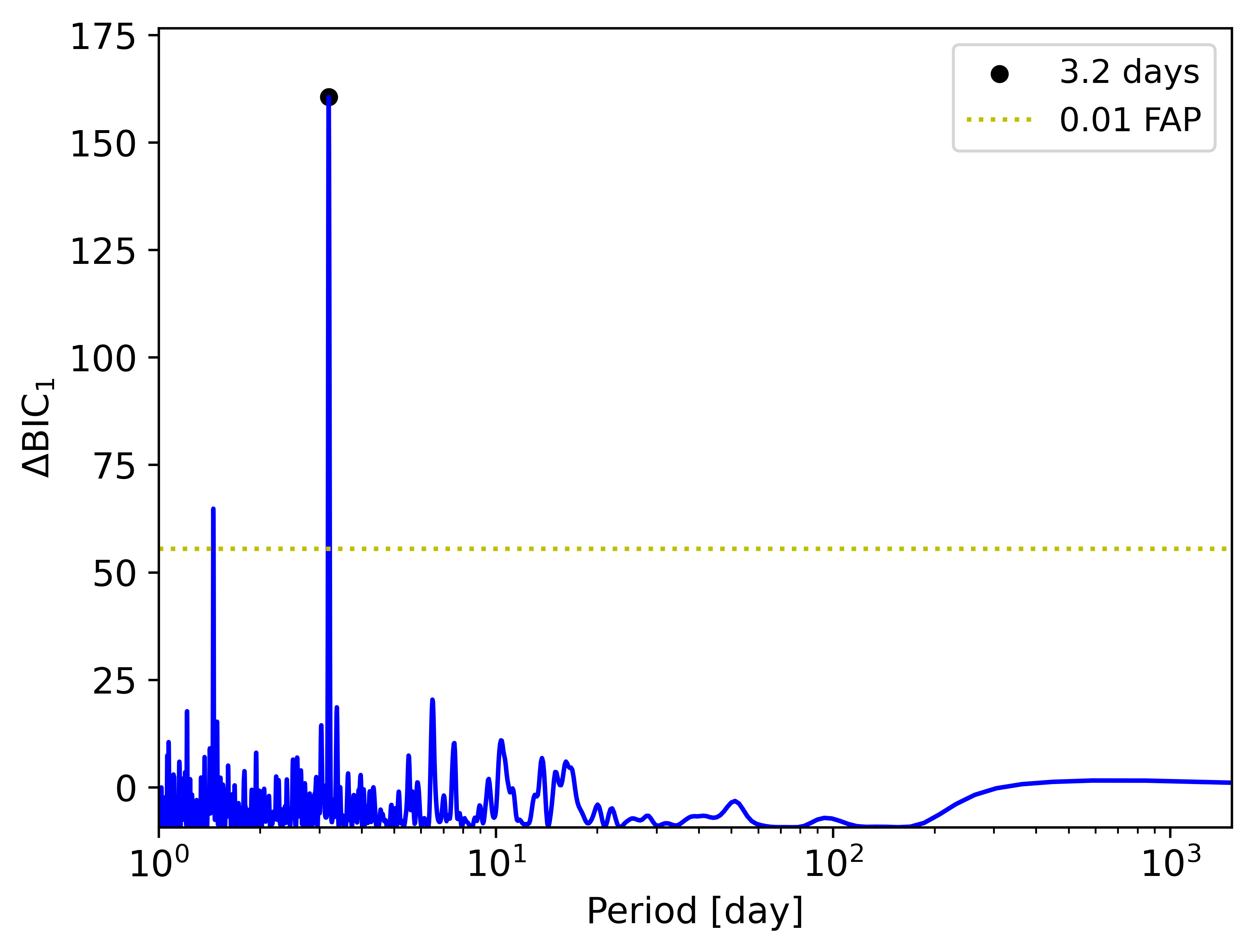}
        \includegraphics[width=0.31 \textwidth]{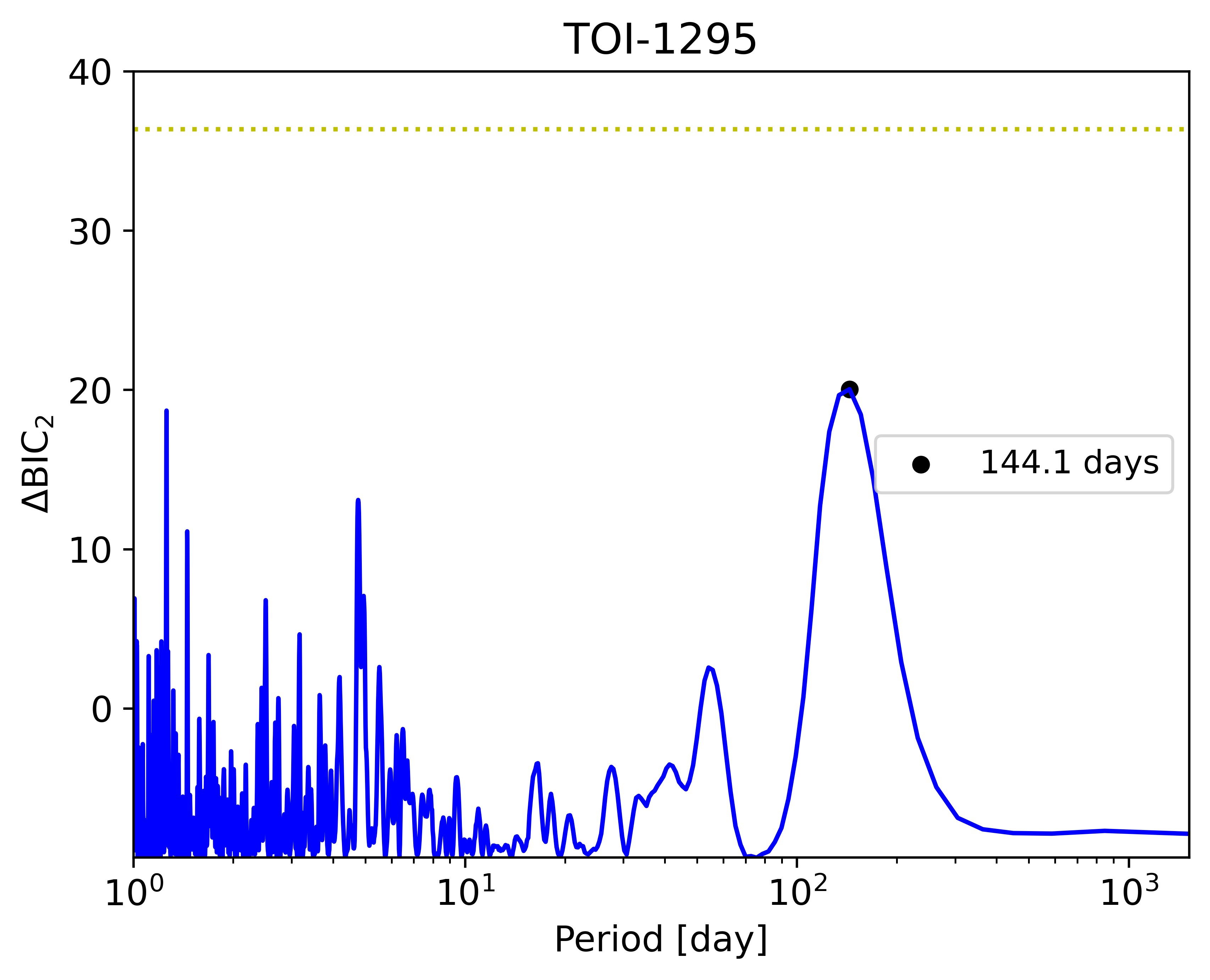}
        \includegraphics[width=0.34 \textwidth]{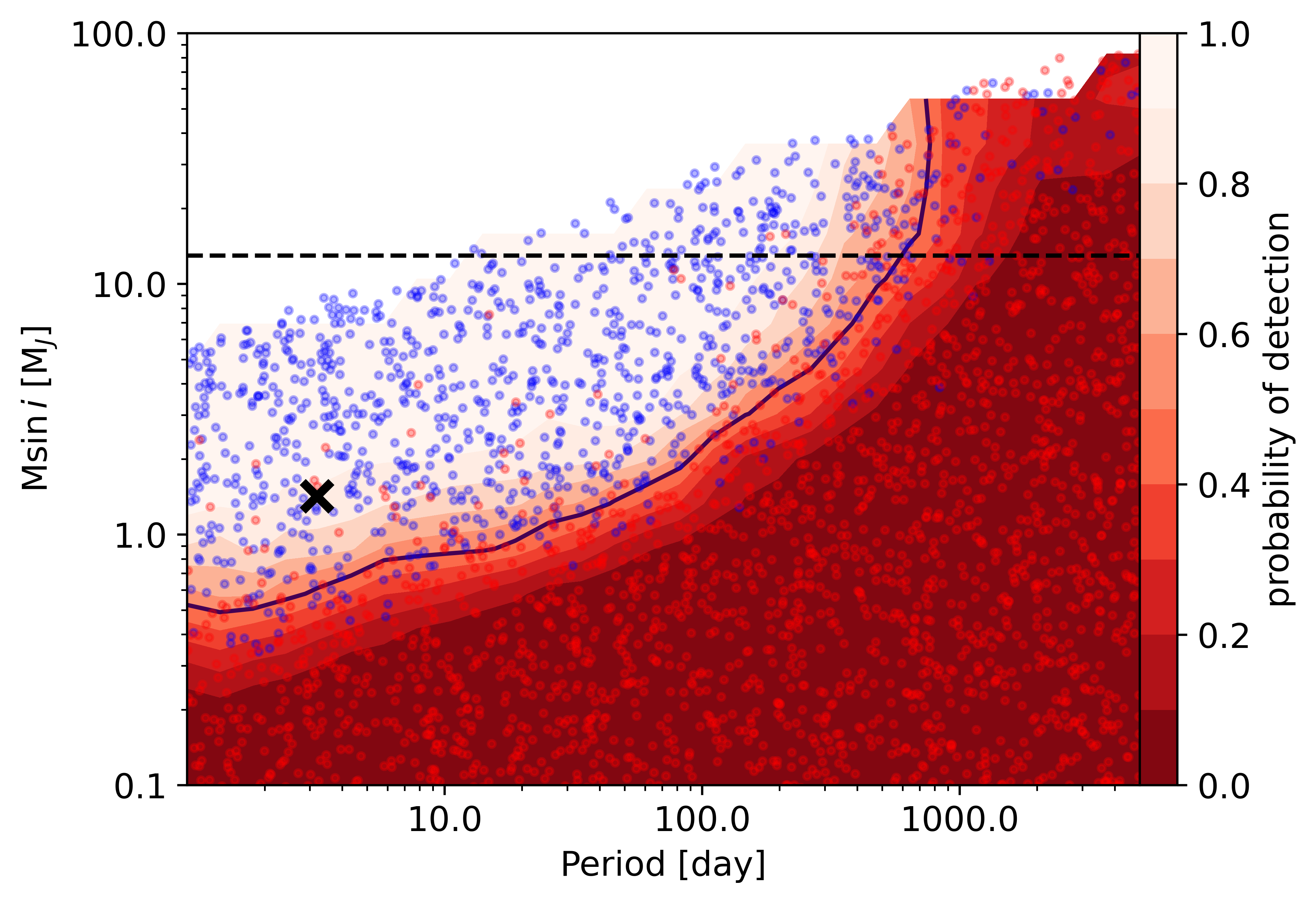}
    \end{minipage}
    \centering
    \begin{minipage}{\textwidth}
        \includegraphics[width=0.31 \textwidth]{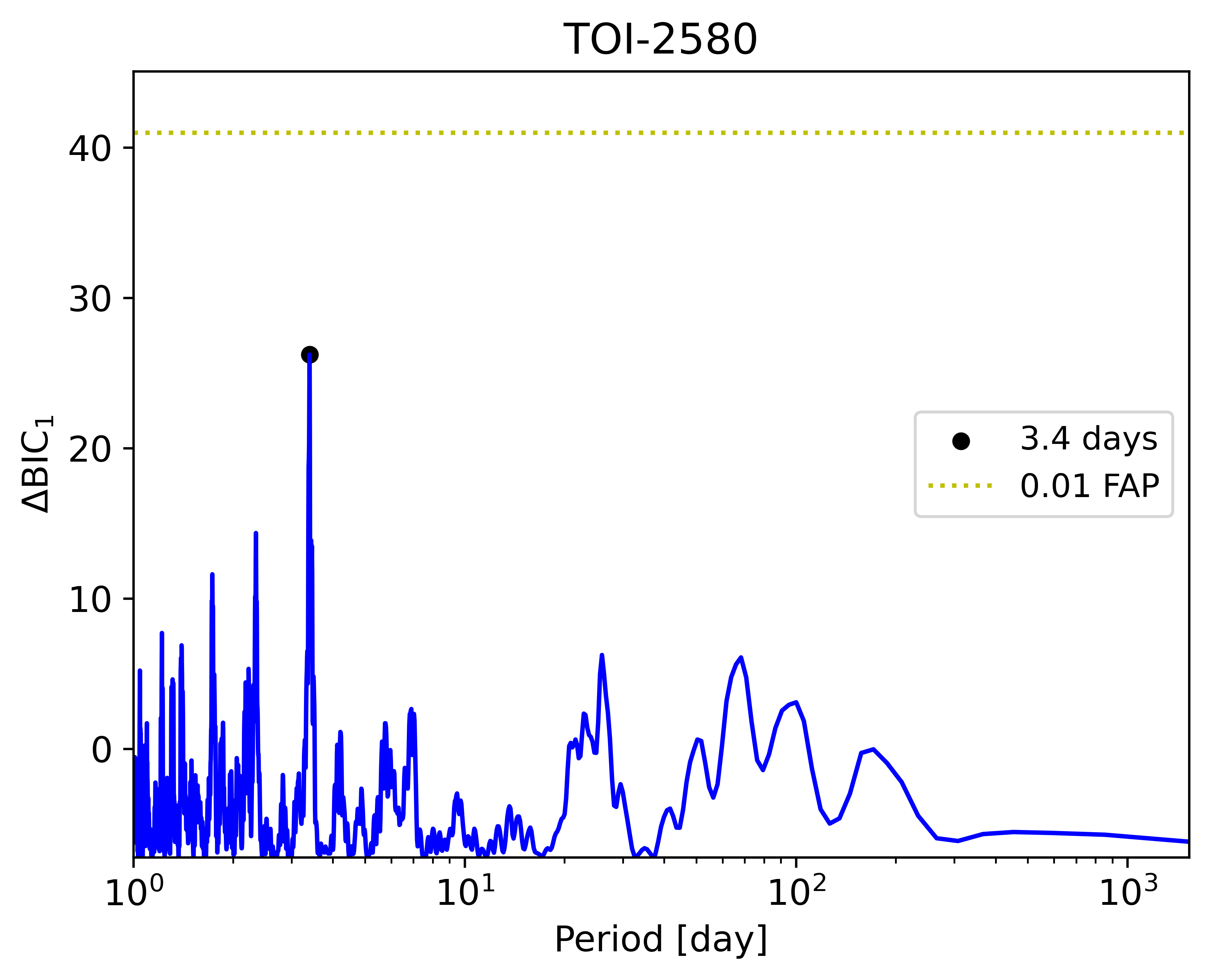}
    \end{minipage}
    \centering
    \begin{minipage}{\textwidth}
        \includegraphics[width=0.31 \textwidth]{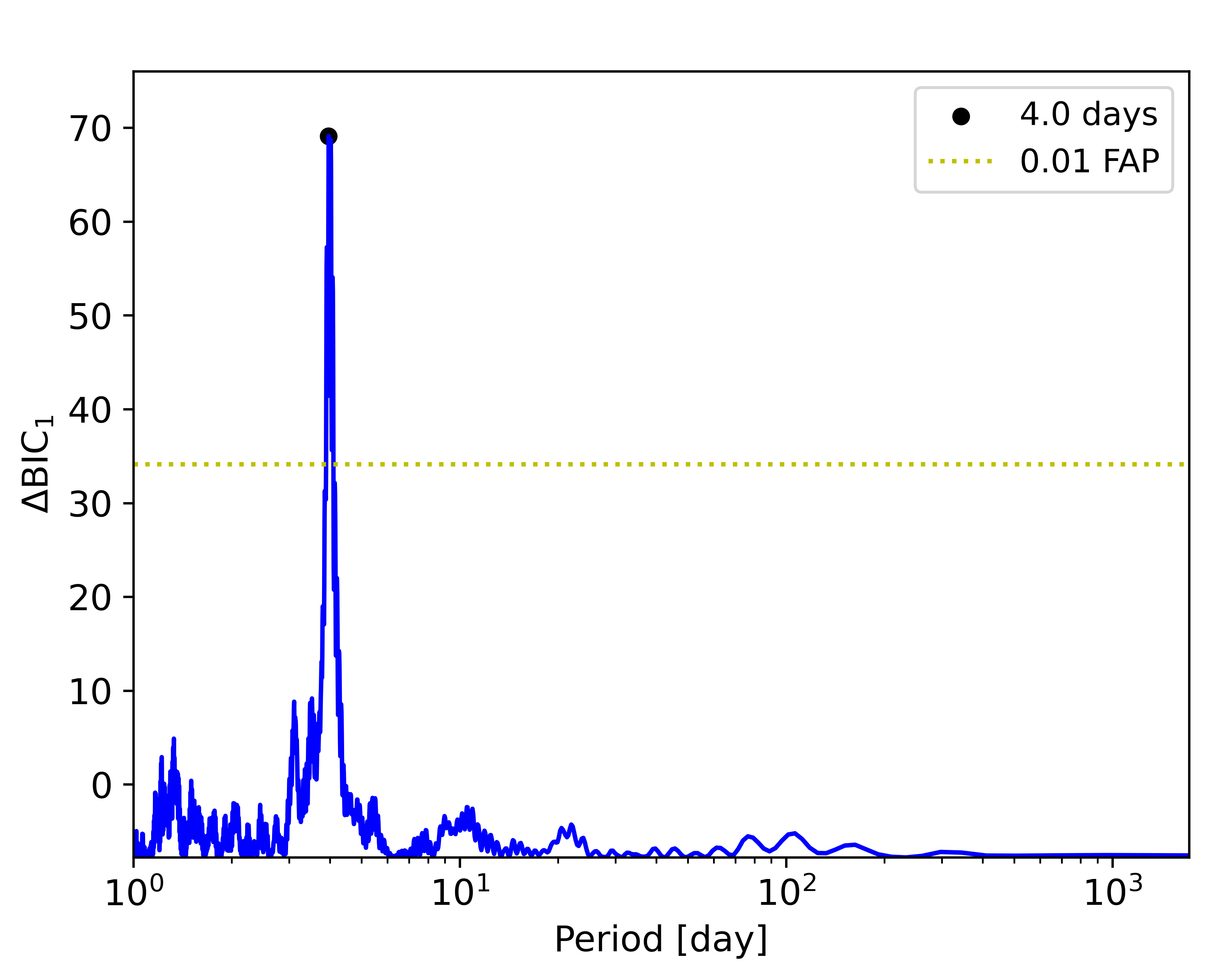}
        \includegraphics[width=0.31 \textwidth]{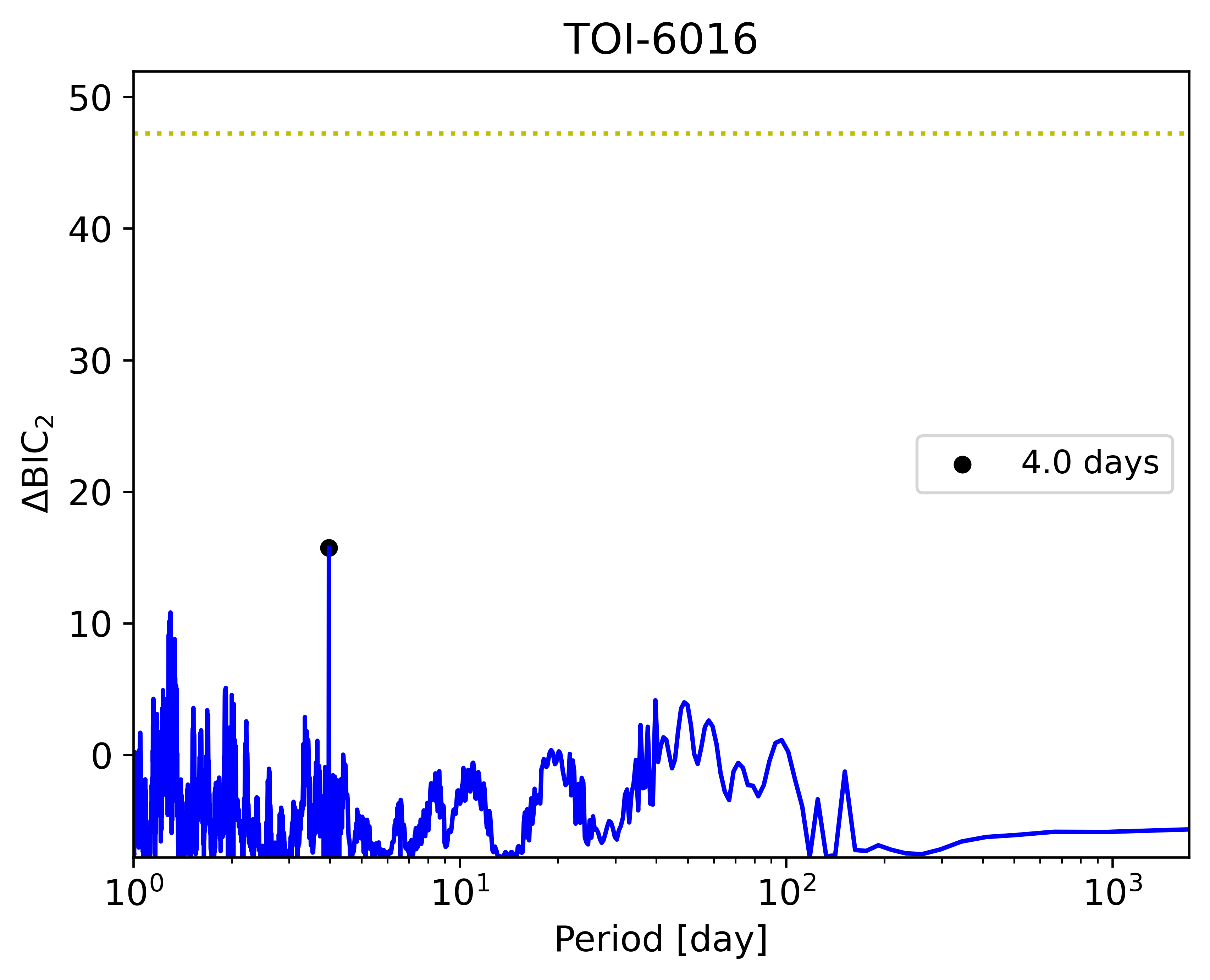}
        \includegraphics[width=0.34 \textwidth]{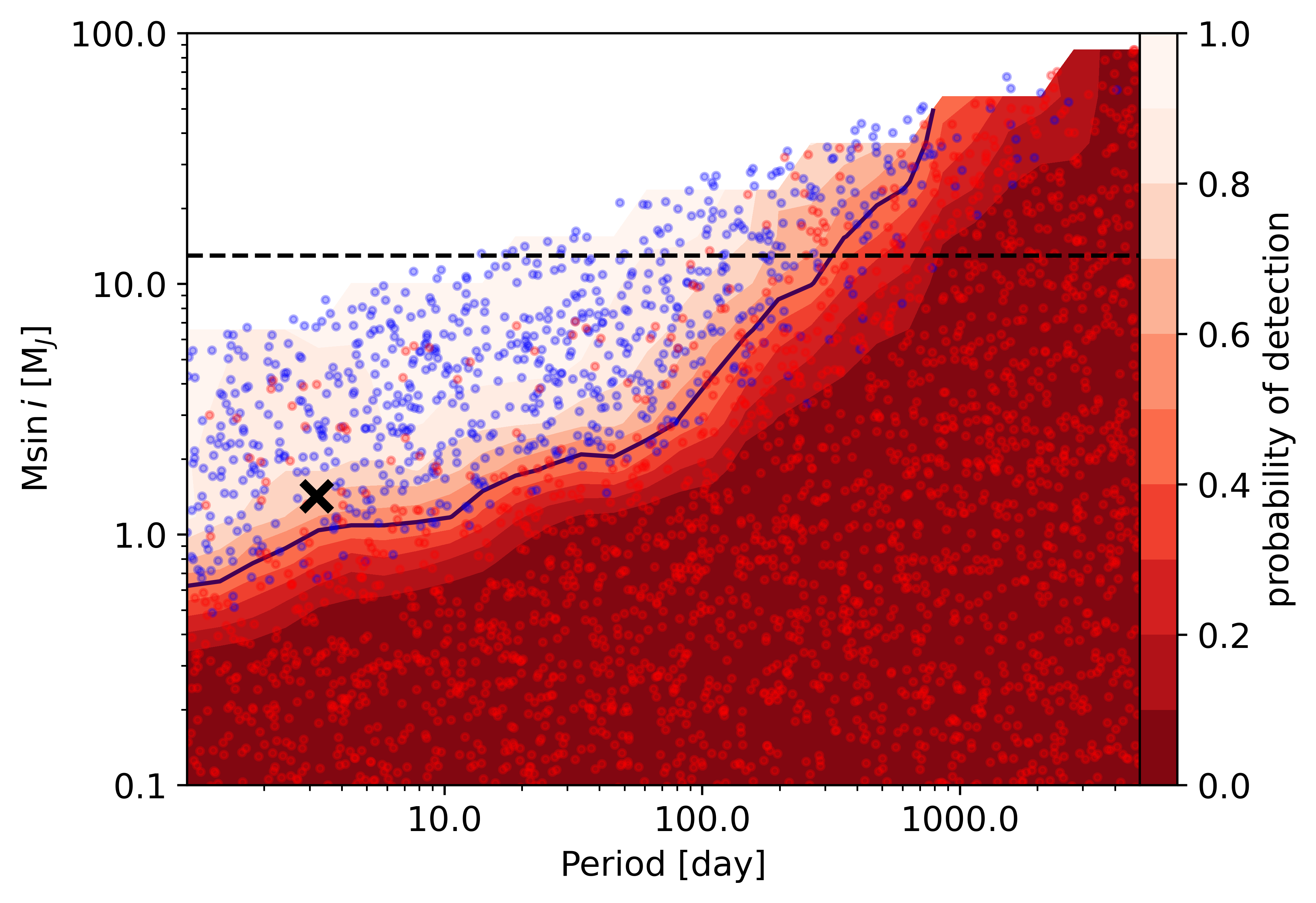}
    \end{minipage}
    \centering
    \begin{minipage}{\textwidth}
        \includegraphics[width=0.31 \textwidth]{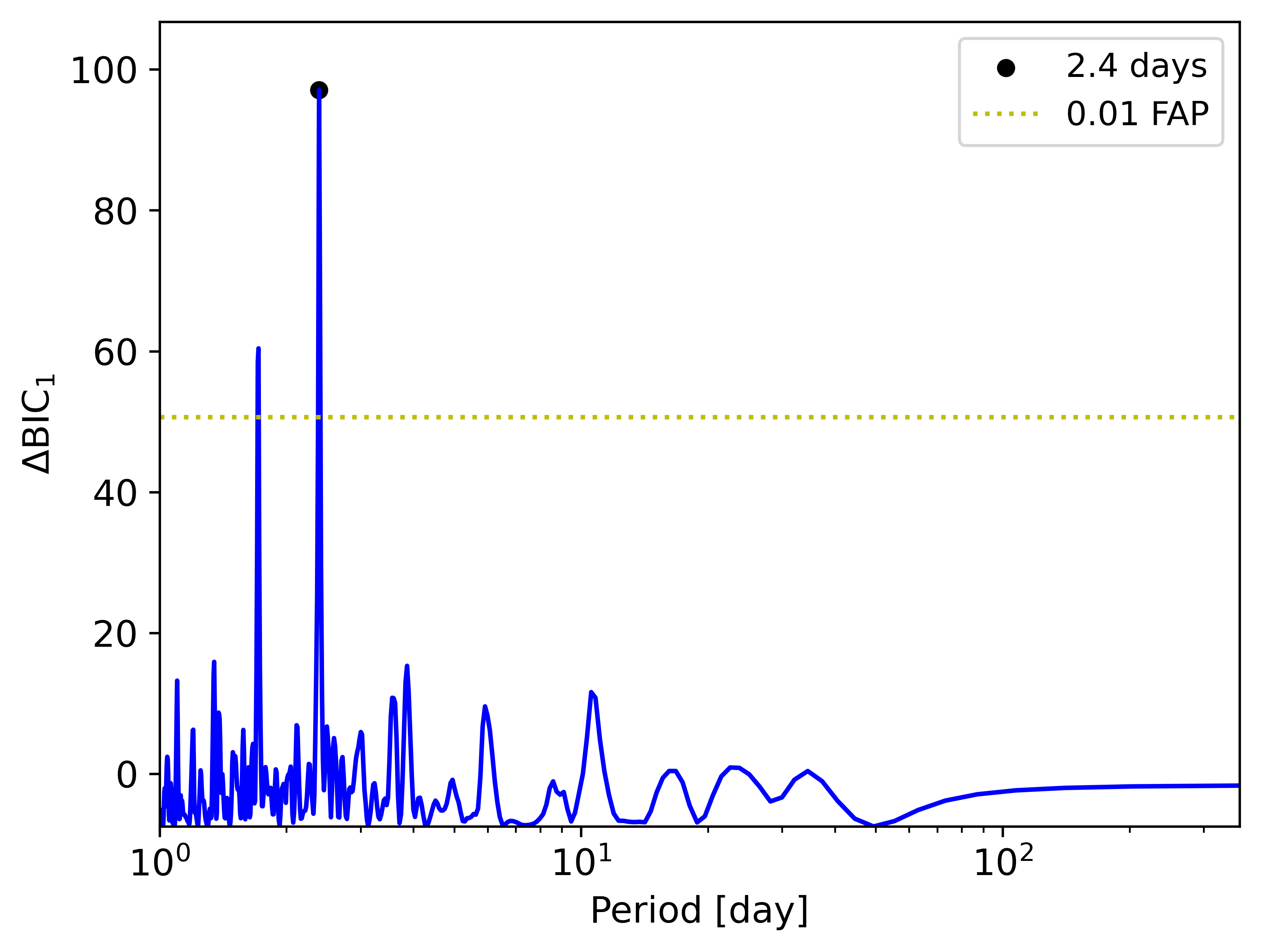}
        \includegraphics[width=0.31 \textwidth]{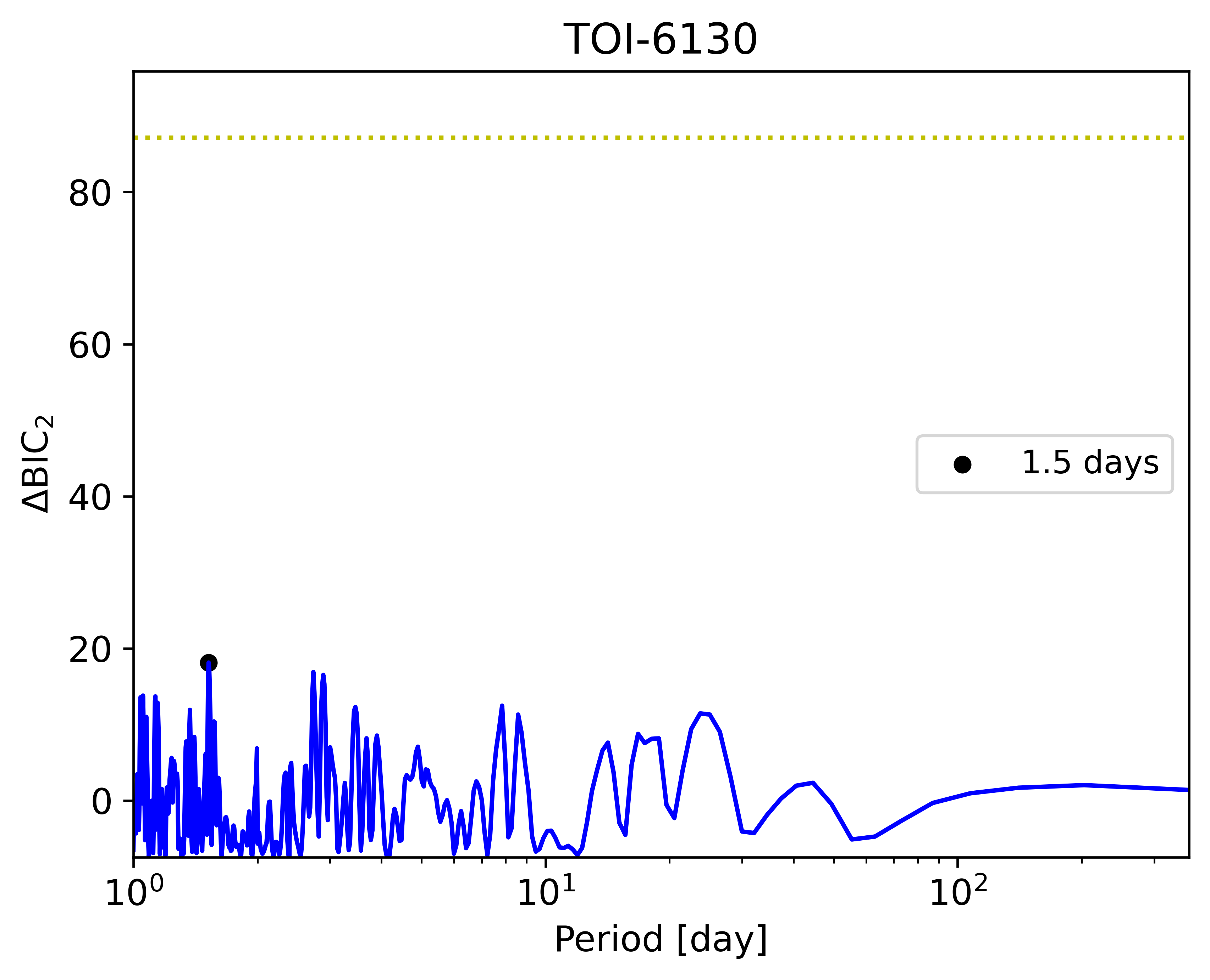}
        \includegraphics[width=0.34 \textwidth]{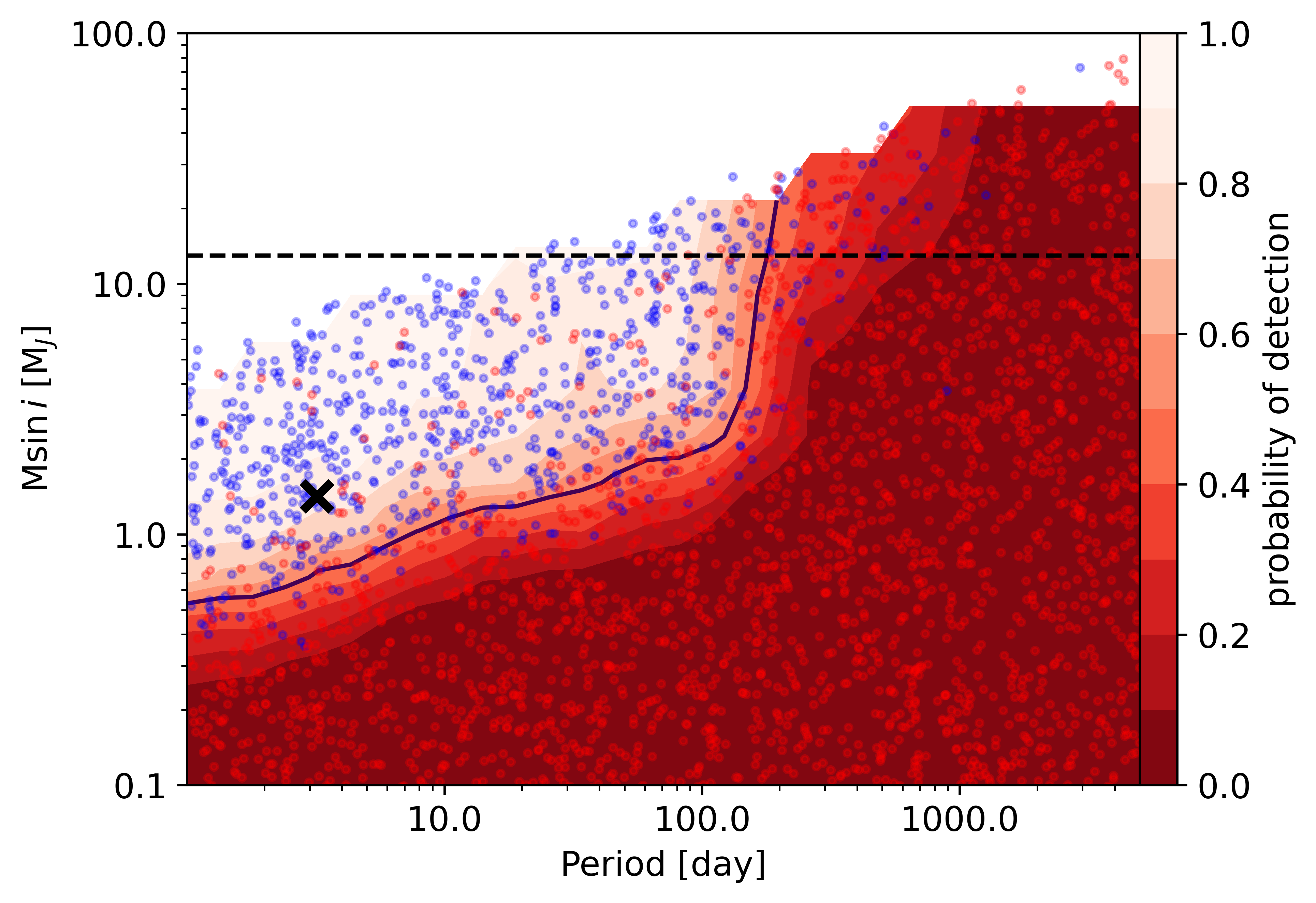}
    \end{minipage}
    \caption{Results from RV-search for TOI-1295, TOI-2580, TOI-6016, and TOI-6130 (from top to bottom). Left: Periodogram for one planet retrieved using the RVs. Middle: Periodogram for a two-planet model with the parameterization for one planet determined with the results from the left panel. Right: Completeness plots: With the available RV data, the blue dots represent the planets that could be recovered, while the available data are not sufficient to detect the planets marked with red circles.}
    \label{fig:RVsearch}
\end{figure*}

\begin{figure}[htp]
    \centering
    \includegraphics[width=0.45\textwidth]{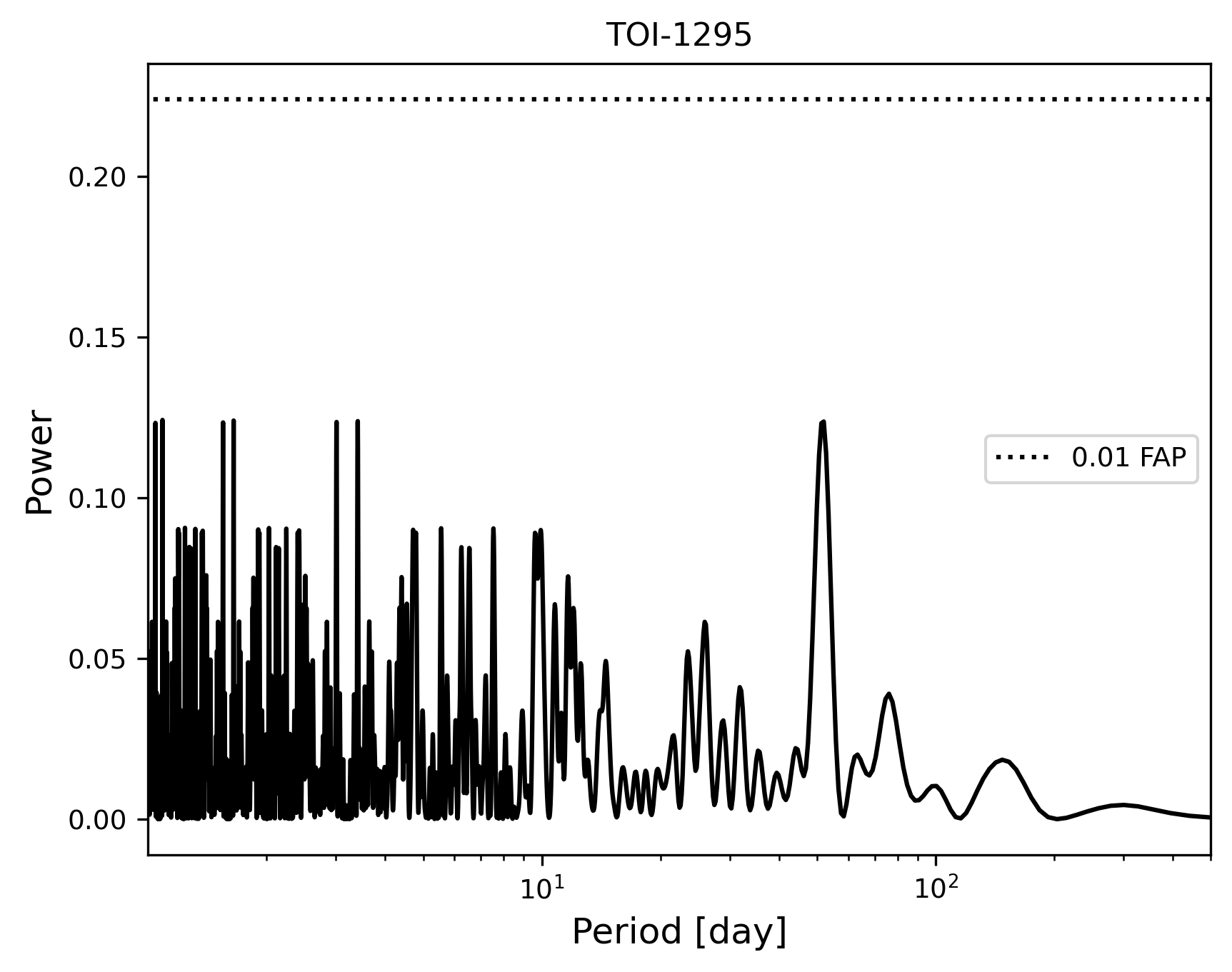}
    \hfill
    \centering
    \includegraphics[width=0.45\textwidth]{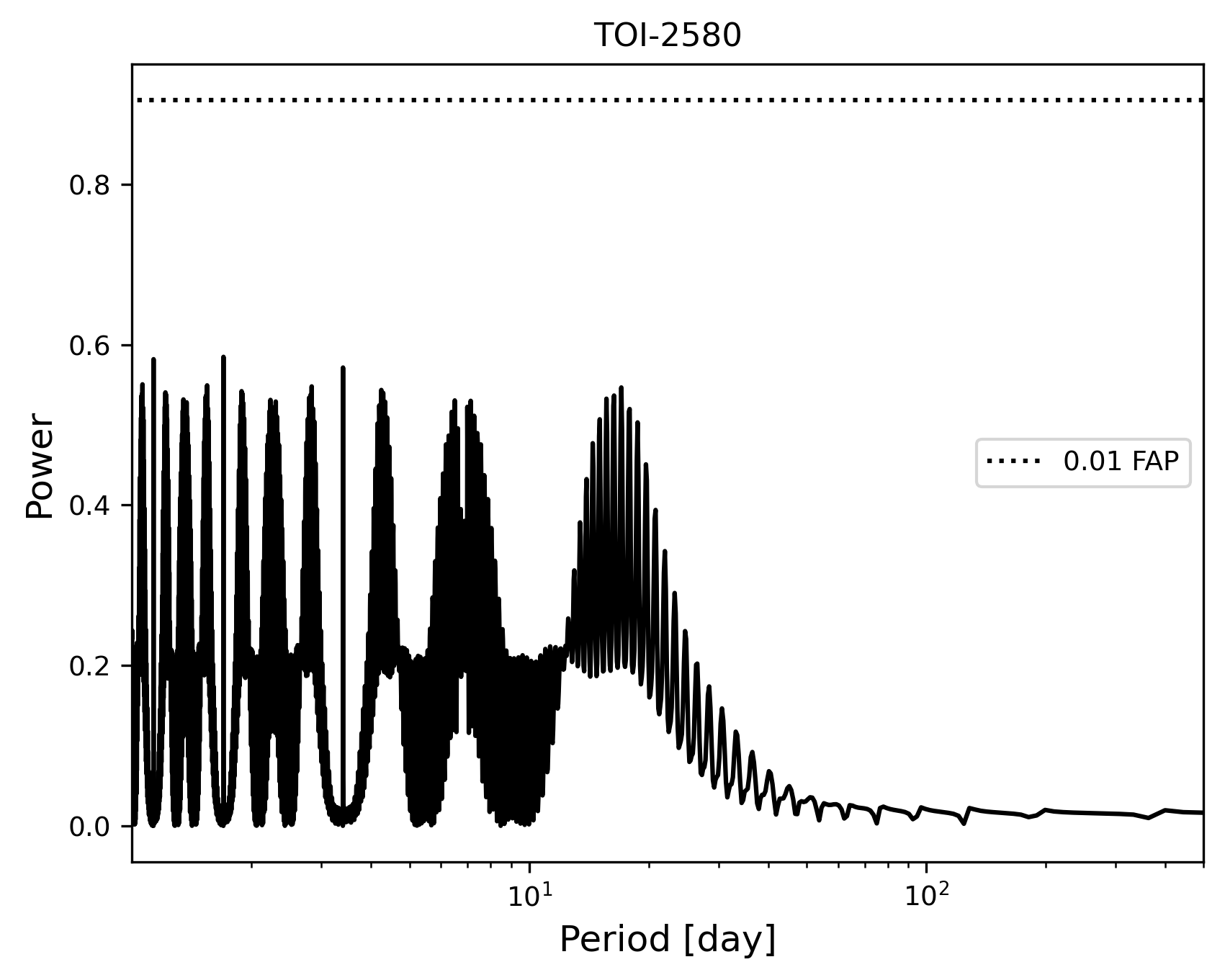}
    \hfill
    \centering
    \includegraphics[width=0.45\textwidth]{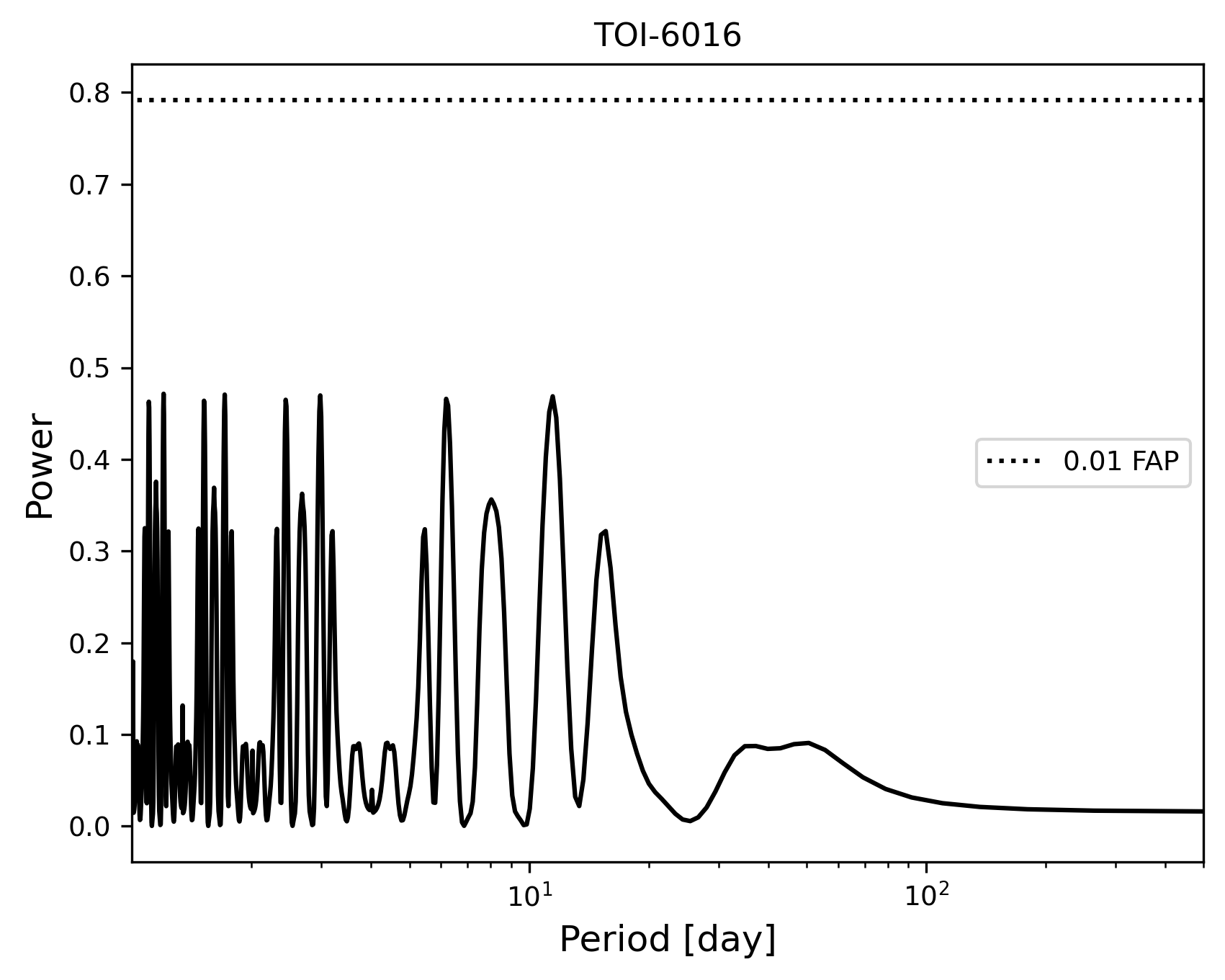}
    \hfill
    \centering
    \includegraphics[width=0.45\textwidth]{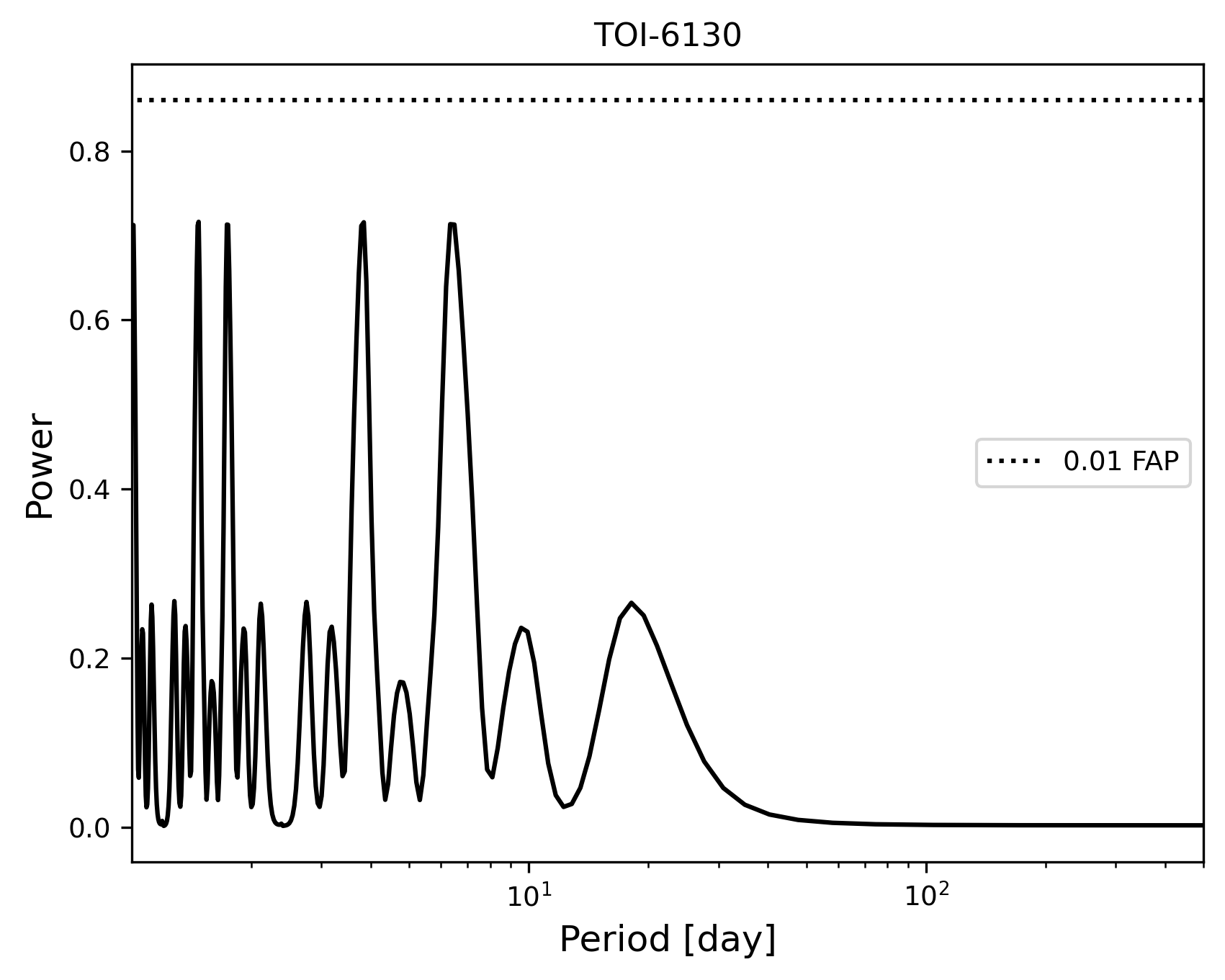}
    \caption{Lomb-Scargle periodograms of the TTVs for our four targets. The dotted line shows the analytical false-alarm probability of 1 \%.}
    \label{fig:LS}
\end{figure}

\end{appendix}

\end{document}